\documentclass[3p]{elsarticle}

\usepackage{setspace} 

\usepackage{textcomp}	 

\usepackage{setspace}
\usepackage{booktabs}    

\usepackage{amsmath} 	  

\usepackage{url}		

\usepackage{graphicx}    
\usepackage{epstopdf}    

\usepackage{subfigure}	 

\journal{Journal of The Electrochemical Society}

\begin{document}

\begin{frontmatter}

\title{Using a dual plasma process to produce cobalt--polypyrrole catalysts for the oxygen reduction reaction in fuel cells -- part II: analysing the chemical structure of the films}

\author[INP]{Christian Walter\corref{cor1}}
\ead{walter@inp-greifswald.de}
\address{phone: +49 3834 554416
fax : +49 3834 554301}

\author[ESRF]{Kurt Kummer}

\author[TUD]{Denis Vyalikh}

\author[INP]{Volker Br\"user}

\author[INP]{Antje Quade}

\author[INP]{Klaus-Dieter Weltmann}

\cortext[cor1]{Corresponding author}

\address[INP]{
Leibniz Institute for Plasma Science and Technology,
INP Greifswald e.V.,
Felix-Hausdorff-Str. 2,
17489 Greifswald,
Germany
}

\address[ESRF]
{European Synchrotron Radiation Facility, 
6 Rue Jules Horowitz, 
B.P. 220, 
F-38043 Grenoble Cedex, 
France}

\address[TUD]{
Institut f\"ur Festk\"orperphysik,
TU Dresden, 
Zellescher Weg 16,
01062 Dresden,
Germany
}

\begin{abstract}

The chemical structure of cobalt--polypyrrole -- produced by a dual plasma process -- is analysed by means of X-ray photoelectron spectroscopy (XPS), near edge X-ray absorption spectroscopy (NEXAFS), X-ray diffraction (XRD), energy-dispersive X-Ray spectroscopy (EDX) and extended x-ray absorption spectroscopy (EXAFS).It is shown that only nanoparticles of a size of 3\,nm with the low temperature crystal structure of cobalt are present within the compound. Besides that, cobalt--nitrogen and carbon--oxygen structures are observed. Furthermore, more and more cobalt--nitrogen structures are produced when increasing the magnetron power. Linking the information on the chemical structure to the results about the catalytic activity of the films  -- which are presented in part I of this contribution -- it is concluded that the cobalt--nitrogen structures are the probable catalytically active sites. The cobalt--nitrogen bond length is calculated as 2.09\,\AA\ and the carbon--nitrogen bond length as 1.38\,\AA.

\end{abstract}


\end{frontmatter}

%

\section{Introduction}

At present, much effort is being spent on the development of an efficient and economical catalyst for the oxygen reduction reaction in polymer electrolyte membrane fuel cells, and non-noble metal catalysts (NNMCs) have been recognised as a promising alternative to the costly platinum employed to date. 

Plasma depositing techniques are known to be one way of producing various metal--polymer films -- e.g. silver nanoparticles distributed in various matrices \cite{Korner2010,Korner2011,Balazs2007}, Zn:Si:O and Ti:Si:O \cite{Daniel2007}, gold nanoparticles distributed in a fluorocarbon matrix\cite{Despax1989} as well as copper-polypyrrole compounds \cite{Walter2009}. In part I of this contribution results from several electrochemical and surface porosity measurements have been presented showing that NNMCs can be obtained by using a dual plasma-enhanced vapour deposition/plasma vapour deposition process, too \cite{PartI}.

There is an ongoing discussion about chemically active sites in NNMCs, which are produced via pyrolysis -- metal--N$_2$ as well as metal--N$_4$ structures were identified and are thought to be responsible for the catalytic activity of the compounds. \cite{Zhang2008b,Bogdanoff2004,Jaouen2009} The binding state of the nitrogen itself -- pyrollic, pyridinic or edge plane -- may play a role as well. \cite{Faubert1999,Maldonado2005} In recent years it has been claimed that some carbon-based materials that do not contain any transition metals show catalytic activity for the oxygen reduction reaction as well. \cite{Matter2006,Niwa2009,Ozaki2006} This led to the discussion of whether the metal itself is actually part of the catalytic site or if graphitic nitrogen plays the major role while the metal only catalyses the formation of those structures during the preparation process. Naturally, as for the chemically produced NNMCs, the question arises of which site is responsible for the catalytic activity of the NNMCs produced in the dual plasma process. 

Here the NNMCs thus produced are analysed by means of various X-ray methods: Energy dispersive X-ray spectroscopy (EDX), X-ray diffraction (XRD), X-ray photoelectron spectroscopy (XPS), near edge X-ray absorption fine structure spectroscopy (NEXAFS), and extended X-ray absorption fine structure spectroscopy (EXAFS). The obtained results made it possible to draw conclusions on their chemical structures. Combining them with the results of the electrochemical measurements presented in part I, chemical sites were identified that are probably responsible for the catalytic activity of the films.

\section{Materials and Methods}

The experimental setup for the production of the compounds is given in part I of this paper. \cite{PartI}

Energy-dispersive X-Ray spectroscopy (EDX) was carried out at the field emission scanning electron microscope JEOL JSM-7500F with a Bruker X-Flash spectrometer (30\,mm$^2$ silicon drift droplet detector) having an energy resolution of 127\,eV. For the optimal count rate during the EDX analysis the acceleration voltage of the electron beam was adjusted to 5\,kV leading to an information depth of up to about 300\,nm. For these measurements, the films were deposited on a silicon wafer to avoid additional signals from a substrate containing carbon or oxygen.

The phase composition was quantified by X-ray diffraction (XRD) using Cu K$\alpha$ radiation (full pattern method for quantitative phase analysis, HZG4 Seifert FPM). For those measurements, the cobalt-polypyrrole (Co--PPy) films were deposited on undoped silicon wafers in the (100) direction and measured in grazing incidence setup at an angle of 0.5\textdegree\ and 1\textdegree\ to the sample surface.

With X-ray photoelectron spectroscopy (XPS, Kratos analytical Axis Ultra) the content of different chemical elements on the surface (without hydrogen) and their binding composition were measured. The XPS spectra were recorded using a conventional hemispherical analyser and a monochromatic Al K$\alpha$ source at 1486.6 eV with 150\,W X-ray power in a standard configuration. Data acquisition parameters for photoelectron spectra were 80\,eV pass energy for measurements of the element content and 10\,eV for high-resolved peaks. Charge compensation was applied. Data acquisition and processing was realised with Vision 2.1.3 software (operating software: Kratos, Manchester, UK). 
The peak fitting procedure was carried out with the help of CasaXPS software version 2.2 (Casa software Ltd., Teignmounth, UK). Gauss-Lorentz (30\,\% Lorentz) distribution, linear background and fixed full width at half maximum (FWHM) of max. 1.2\,eV were applied. All values are given in XPS atomic percentage and ratios thereof. The XPS atomic percentage is calculated from the measured peak areas for each element present in the spectrum considering their 
respective sensitivity factors. Survey scans and highly resolved spectra of the C, N, O $1s$ and Co $2p$ regions were recorded for the films deposited on silicon wafers (orientation [100]).

NEXAFS experiments at the C, N and O $K$ edge and the Co $L_{2,3}$ edges were performed at the Russian-German beamline (RGBL) of the BESSY II synchrotron facility. To this end, the cobalt-polypyrrole films were deposited onto silicon wafers. All spectra were recorded at normal incidence to the sample surface, using a Keithley source meter to measure the drain current into the sample (total electron yield mode). The probing depth in total electron yield mode at these energies amounts to $\sim10\,$nm. The energy range of 30\,eV up to 1500\,eV can be recorded with a resolving power $E/\Delta E$ of up to 20000. The signal was normalised to both the storage ring current and the energy-dependent photon flux of the RGBL, which was measured using a freshly sputtered gold sample. The background originating from lower lying absorption edges was eliminated by fitting a 2nd order polynomial background to the pre-edge region.

Near edge X-ray absorption fine structure (NEXAFS) and extended X-ray absorption fine structure (EXAFS) of the cobalt $K$ edge was carried out at the beamline KMC-2 at the BESSY synchrotron. There the energy can be tuned from 5\,keV to 14\,keV with a resolving power $E/\Delta E$ of approx. 5000. For those measurements the cobalt-polypyrrole films were deposited onto a teflon substrate to avoid additional signals from heavier elements. It was measured with the fluorescence method using a PIN photodiode with an angle of 45\textdegree\ to the sample. Due to the high penetration depth of X-rays of energies at about 8\,keV these measurements give information on the bulk structure of the compounds. Data analysis was carried out with IFEFFIT version 1.2.11. \cite{IFEFFIT}

\section{Results and Discussion}

\subsection{Energy-Dispersive X-Ray Spectroscopy}

The percentages of the elements detected in the cobalt-polypyrrole films are shown in table \ref{tab:EDX}. It can easily be seen that the cobalt content rises with increasing magnetron power, while nitrogen and carbon content decreases. The oxygen content does not show a clear trend. Because oxygen was not introduced into the plasma process, the oxygen content is likely to come from the oxidation of the cobalt-polypyrrole compounds after the process since the samples were stored in air.  The rather high oxygen content is also related to the fact, that oxygen is mostly present at the surface of the compounds (c.\,f. XPS section). Since the analysis method of the EDX quantification is sensitive to surface contaminations, this leads to an overestimation of the oxygen content within the samples. \cite{Eggert2005}

Because no further information can be obtained directly from the elemental contents alone, the ratio of nitrogen versus carbon was calculated. These values make it possible to obtain more detailed information about the structural composition of the films.
As can be seen in table \ref{tab:EDX}, this value is 0.25 at pure polypyrrole (which corresponds to the 1 nitrogen atom versus 4 carbon atoms in pure pyrrole) and increases for the cobalt containing samples up to 0.31.

\subsection{X-Ray Diffraction}

\begin{figure}[h]
	\centering
	\includegraphics[width=\linewidth]{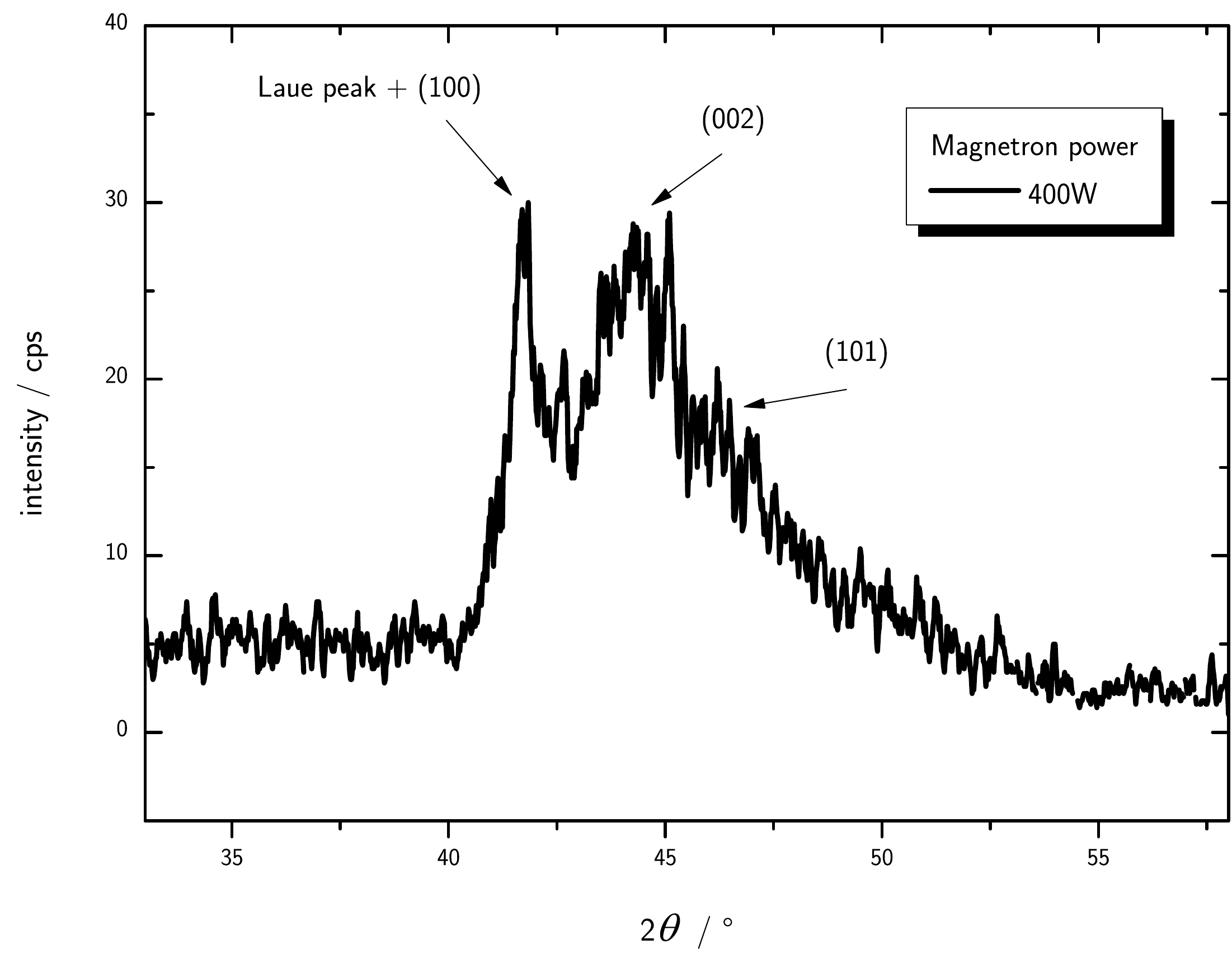} 
	\caption{XRD spectrum of cobalt-polypyrrole produced with a magnetron power of 400\,W; Laue peaks from the silicon substrate and peaks belonging to cobalt nanoparticles with hcp structure are labelled}
	\label{fig:XRD_400W}
\end{figure}

Exemplarily, the XRD spectra of a film produced with a magnetron power of 400\,W is presented in figure \ref{fig:XRD_400W}. Peaks evolving from metallic cobalt and a Laue-peak originating from the silicon substrate can be seen. It is interesting to note that only metallic cobalt in the hcp crystal structure ($\alpha$ cobalt) can be seen -- in contrast to the cobalt with fcc structure ($\beta$ cobalt) that is found at pyrolised samples. \cite{Yang2007a,Garsuch2008,Yuasa2005,Sirk2008} This is because cobalt transforms from the $\alpha$ to the $\beta$ state at 471\,$^\circ$C. \cite{Holleman1995} Therefore, the observation that only the $\alpha$ state of cobalt is found in the plasma-produced samples is a proof that the substrate is kept at low temperatures during the plasma process.

By using the Williamson-Hall-Plot \cite{Williamson1953} it has been calculated that nanoparticles with a grain size of 3\,nm and a lattice spacing of 2.493\,\AA\ are present within the sample. Comparing this to a grain size of about 20\,nm in a pyrolised sample \cite{Yang2007} heated up to 900\,$^\circ$C one can see that -- bearing in mind that the cobalt content is 61\,at.-\% for the sample produced in this contribution compared to about 13\,at.-\% for the heat treated sample -- the clustering is strongly suppressed in the plasma-produced films compared to the pyrolised ones. The reason for that is the fact that the atom mobility is smaller at lower temperatures. This decreased atom mobility could also lead to an increased amount of catalytically active, cobalt containing structures since fewer cobalt atoms might be able to leave the Co--N$_x$-structures during the production process forming catalytically inactive metallic cobalt.

\subsection{X-Ray Photoelectron Spectroscopy}

The elemental content calculated from XPS-measurements is given in table \ref{tab:XPS}.
Since the elemental contents measured with XPS are always contaminated with atmospheric hydrocarbons -- the depth of information is only up to about 10\,nm -- the calculation of the N/C ration as done in the EDX section does not lead to meaningful data here. Furthermore, the nitrogen content at the surface is at the detection limit of the XPS measurements -- the calculated N/C value would be very small and the error level high. This leads to the observation that the error level of the N/C calculation would be higher than the N/C value itself -- so it is not valuable at all. But qualitatively the same conclusion can be drawn from the elemental contents measured by XPS. The cobalt content rises with increasing magnetron power while nitrogen and carbon content decreases. The oxygen content is much higher in the cobalt-polypyrrole samples than in the polypyrrole itself and increases slightly with increasing power.

By comparing the elemental contents obtained from EDX with the ones measured with XPS one can see that the percentage of carbon and oxygen is higher in the XPS data and the one of nitrogen and cobalt is lower. This is a proof for the assumption from the EDX section that oxygen is only present at the surface of the films. Because of the the much lower information depth of XPS (about 10\,nm) to EDX (about 300\,nm) an increase of carbon and oxygen content by atmospheric contaminations can be seen in the XPS data. \cite{Vickerman1997}

\begin{figure}[h]
	\centering
	\includegraphics[width=\linewidth]{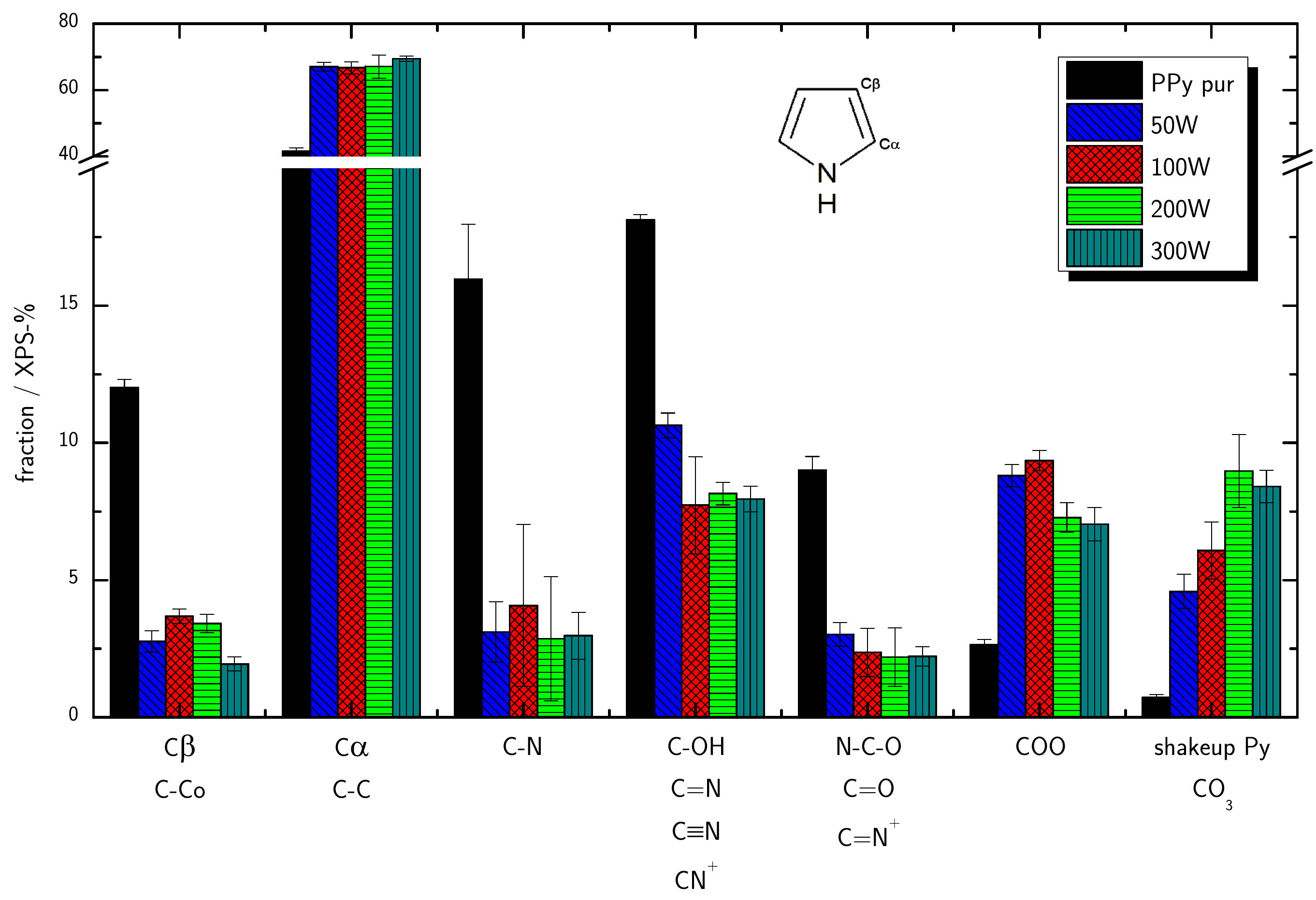} 
	\caption{Summary of the binding type contents calculated from the C $1s$ highly resolved spectra obtained by XPS measurements; the given
power values represent the magnetron power, the RF plasma source was operated at 40\,W for all samples}
	\label{fig:XPS_contents}
\end{figure}

The C $1s$ high resolution spectra were analysed with the binding energies given in table \ref{tab:XPS_bindingenergy}. The results can be seen in figure \ref{fig:XPS_contents}. The binding energy of C$_\alpha$ is different to C$_\beta$ because of the electronegativity of the nitrogen atom within pyrrole. (c.\,f. chemical structure of pyrrol in figure~\ref{fig:XPS_contents}). \cite{Pfluger1984}

It has to be noted that throughout this contribution by using the term \emph{pure polypyrrole} we are referring to plasma-produced polypyrrole that was produced by only using the RF-electrode as explained in part I. \cite{PartI} This is important because by using a plasma process to polymerise pyrrole, a fraction of pyrrole rings is destroyed and various bonds such as e.g. C=N, C=O, N--C--O, COO etc. -- that would not be present in perfect polypyrrole -- are formed. \cite{Zhang1997a} This can also be seen by the fact that there is more C$_\alpha$ then C$_\beta$ in plasma polymerised polypyrrole due to structural disorder -- in perfect polypyrrole both C$_\beta$ and C$_\alpha$ should be equally distributed. In chemically produced polypyrrole, also 1/3 of all pyrrole rings are affected by structural disorder leading to a decrease of the C$_\beta$ peak. \cite{Pfluger1984} However, because of the reactions in the plasma chamber as well as the increased energy input to the substrate the structural disorder in plasma polymerised compounds is even higher. Another reason for the increase of the C$_\alpha$-peak is, that it is overlapping with contaminations of hydrocarbons, that are present in all XPS-spectra.

When comparing the pure polypyrrole samples with the ones containing cobalt one can see that only crosslinked carbon bonds (C--C) and oxidised carbon structures increase while all other bindings decrease.  This could already be seen at copper-polypyrrole samples produced with a dual plasma process \cite{Walter2009} and can be linked to an increase of the energy input into the substrate with rising magnetron power leading to the destruction of the pyrrole rings -- and thus the production of more and more unsaturated bonds that are then oxidised when brought in contact with air. The peak of carbon--cobalt complexes is overlapping with the one of C$_\beta$ so the formation of Co--C bonds can not completely be excluded. However, if Co--C bonds are present, only a small fraction of carbon is bound in that way.

Furthermore, single bonds (C--N) are decreasing more strongly than double or triple bonds. This has also already been seen in copper-polypyrrole and was linked to the hydrogen deficit in pyrrole. \cite{Walter2009} 

Comparing the cobalt-polypyrrole compounds produced with different magnetron powers no clear trend can be seen -- a structure responsible for the increasing catalytic activity with raising magnetron power cannot be identified. Because of that, further experimental techniques are necessary to obtain a deeper insight into the chemical structure of the films. Furthermore, the high resolution spectra of the nitrogen $1s$ region could not be analysed due to the low nitrogen content at the surface leading to a noisy spectrum -- fitting of the peaks obtained there does not lead to meaningful data. 

Because of this, NEXAFS measurements were carried out at the beamline RGBL at BESSY to resolve the chemical structures that are present within the plasma-produced compounds.

\subsection{Near-Edge X-Ray Absorption Fine Structure}

In order to gain a deeper insight into the chemical bonding and -- in particular -- into the chemical state of nitrogen, X-ray-absorption spectroscopy was carried out and the fine structure close to the characteristic absorption edges was analysed. The near-edge fine structure reflects X-ray induced electron transitions from deeply bound core-levels into unoccupied molecular orbitals close to the Fermi level. Since the molecular orbitals are responsible for the chemical bonding, very direct information on the chemical state of atoms can be obtained with element specificity.

As shown in part I of this contribution, the catalytic activity of the Co-polypyrrole films increases with magnetron power \cite{PartI}. Therefore, we tried to carefully follow changes in the near-edge structures of the C, N, O, Co $K$ edges and the Co $L_{2,3}$ edges as a function of magnetron power in order to identify the chemical sites that are responsible for the catalytic activity. Particularly interesting are chemical structures related to those spectral features that become enhanced or exclusively appear in samples produced with high magnetron powers.

\begin{figure}[h]
	\centering
	\includegraphics[width=\linewidth]{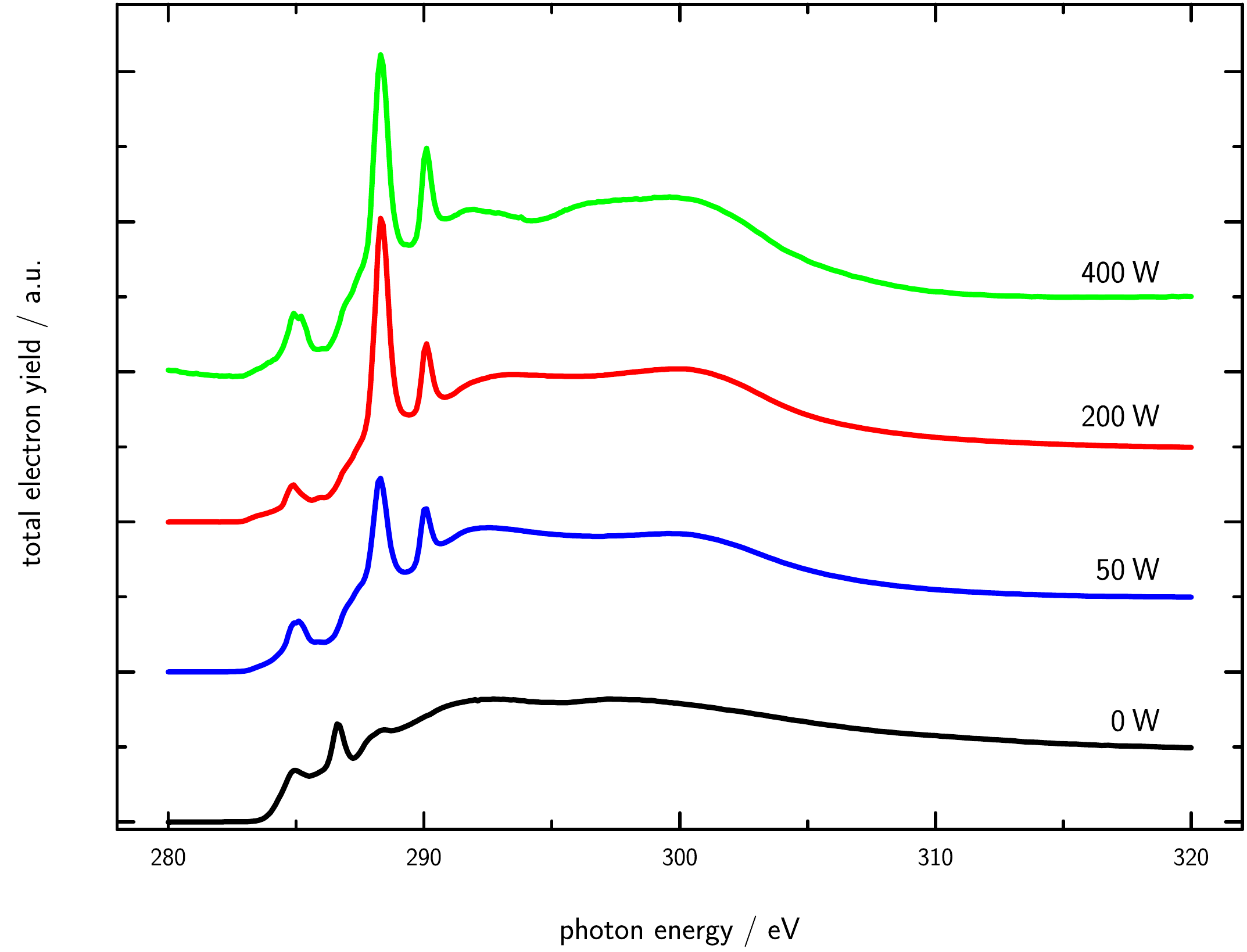} 
	\caption{C $K$ edge spectra of polypyrrole and cobalt-polypyrrole samples produced with different magnetron powers; the assignment of the peaks is given in table \ref{tab:NEXAFSenergies}; the spectra were shifted along the y-axis}
	\label{fig:NEXAFS_C1s}
\end{figure}

\begin{figure}[h]
	\centering
	\includegraphics[width=\linewidth]{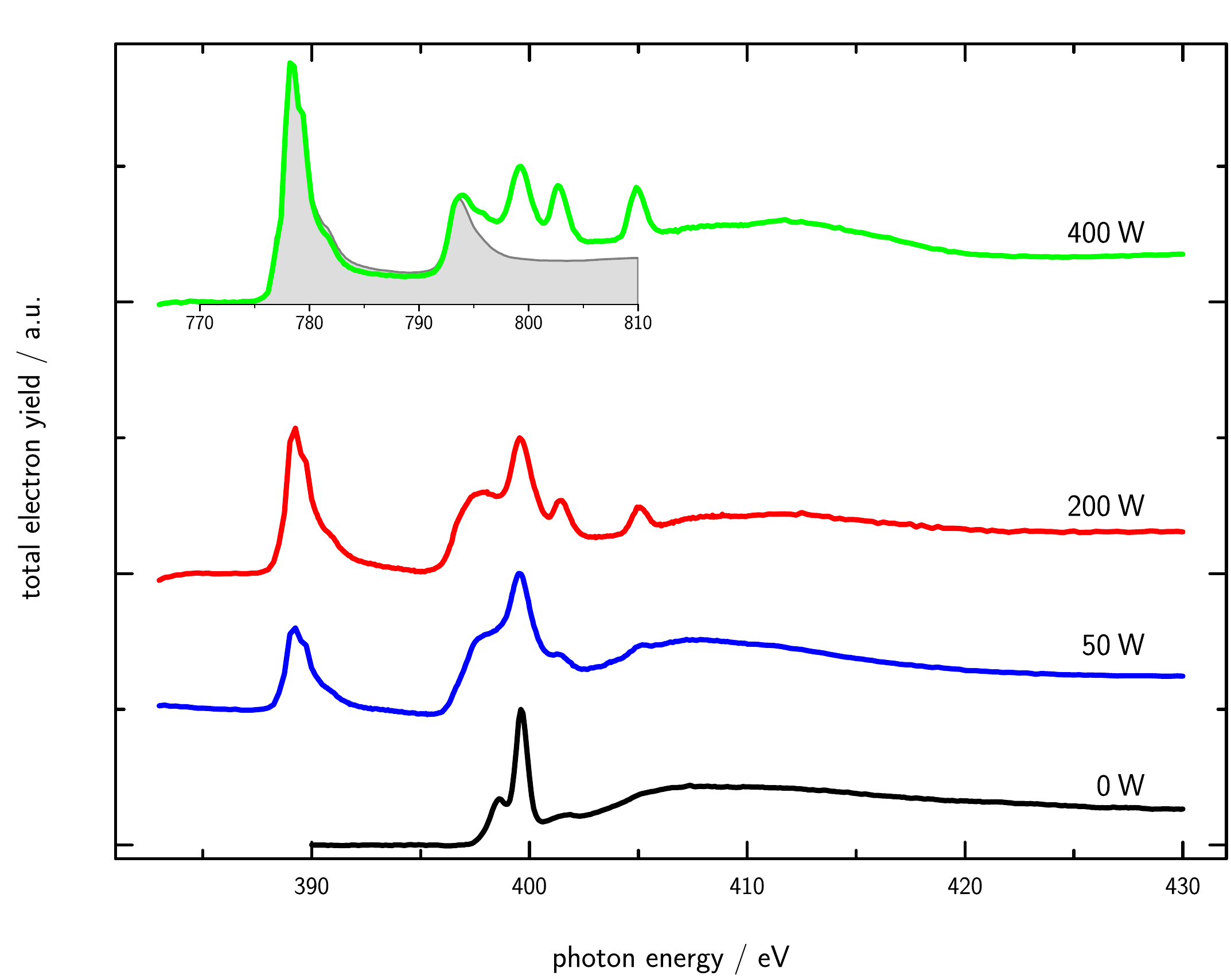} 
	\caption{N $K$ edge spectra of polypyrrole and cobalt-polypyrrole samples produced with different magnetron powers; the assignment of the peaks is given in table \ref{tab:NEXAFSenergies}; the spectra were shifted along the y-axis; the inset shows the cobalt $L_{2,3}$ edges that were measured between 710\,eV--810\,eV and overlap with the nitrogen spectra because of second order light}
	\label{fig:NEXAFS_N1s}
\end{figure}

\begin{figure}[h]
	\centering
	\includegraphics[width=\linewidth]{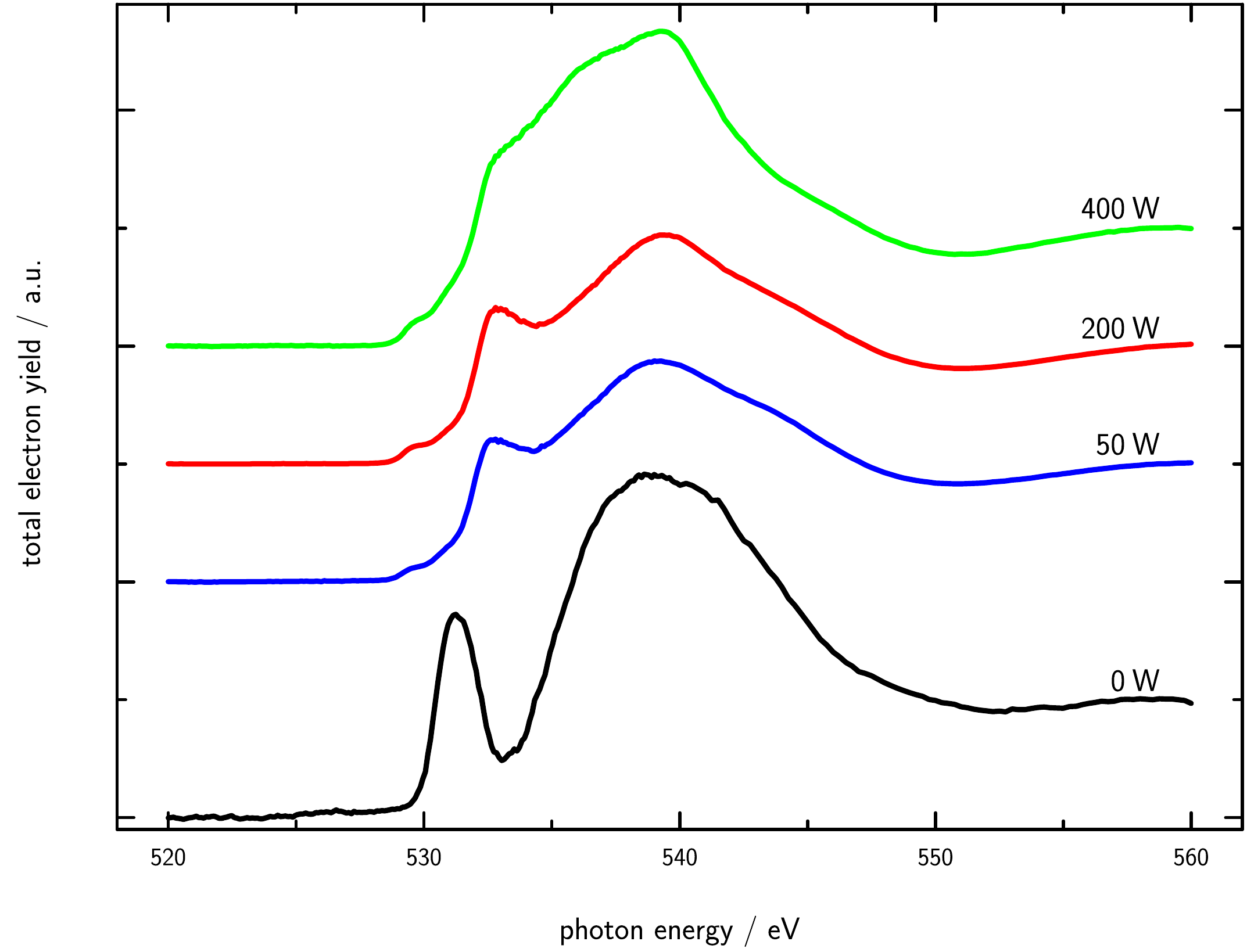} 
	\caption{O $K$ edge spectra of polypyrrole and cobalt-polypyrrole samples produced with different magnetron powers; the assignment of the peaks is given in table \ref{tab:NEXAFSenergies}; the spectra were shifted along the y-axis}
	\label{fig:NEXAFS_O1s}
\end{figure}

\begin{figure}[h]	
	\centering
	\includegraphics[width=\linewidth]{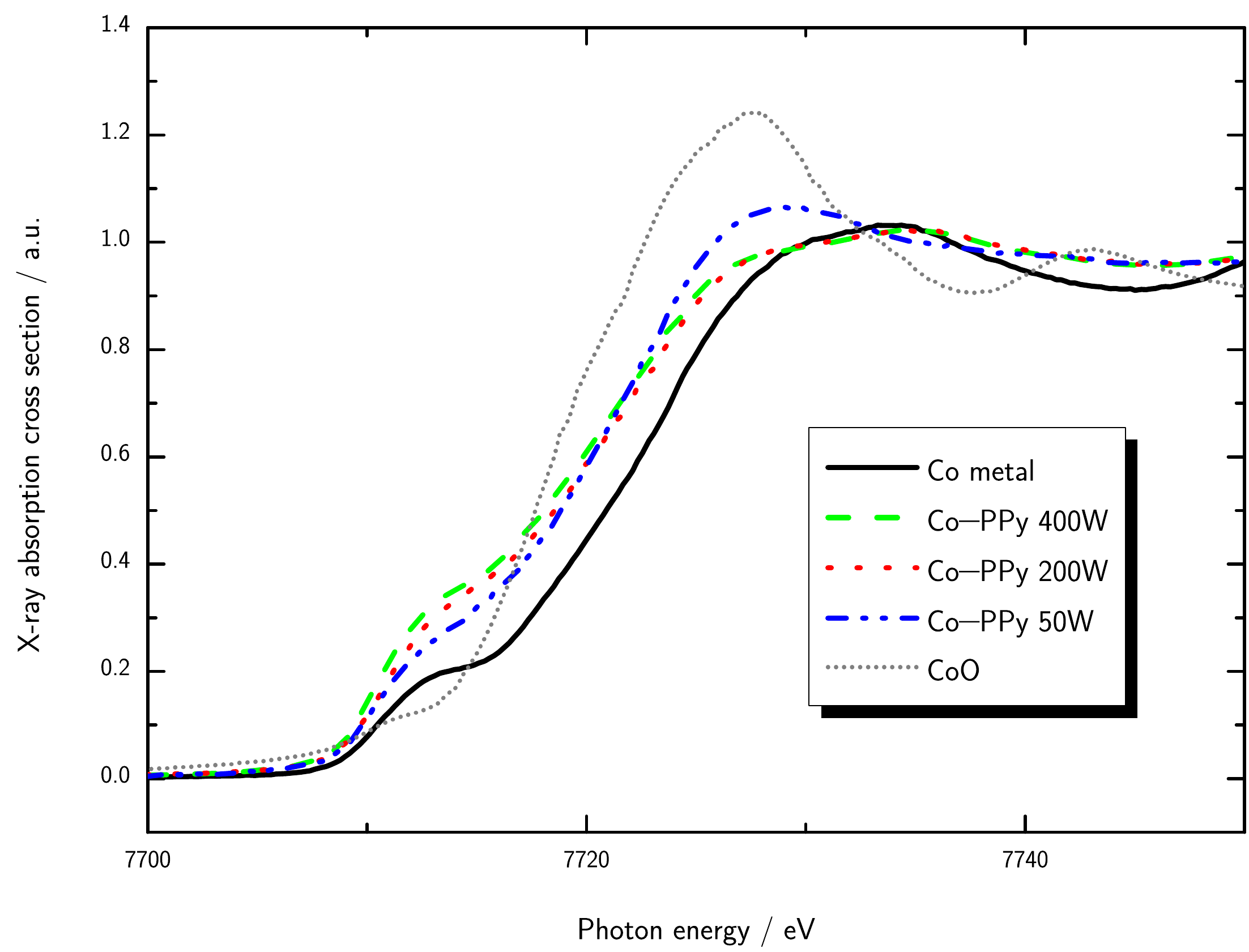} 
	\caption{Co $K$ edge spectra of polypyrrole and cobalt-polypyrrole samples produced with different magnetron powers; the assignment of the peaks is given in table \ref{tab:NEXAFSenergies}}
	\label{fig:NEXAFS_Co1s}
\end{figure}

The spectra recorded at the different absorption edges are shown in figures \ref{fig:NEXAFS_C1s}-\ref{fig:NEXAFS_Co1s}. The assignments of the individual resonances/peaks to certain chemical sites is summarised in table \ref{tab:NEXAFSenergies}. No quantitative fitting was carried out due to the large number of spectral features. The spectra will be discussed only qualitatively, based on the relative height of the peaks and their evolution with magnetron power.

\subsubsection{Carbon $K$ edge}

The carbon $K$ edge spectra are shown in Figure~\ref{fig:NEXAFS_C1s} for a selected set of samples. For native polypyrrole (0\,W magnetron power) the spectrum only consists of peaks which can be attributed to C=C, aromatic C--N structures, and a small contribution that can be assigned to C=O structures.

When the magnetron was used to incorporate cobalt into the films the peak assigned to C=O structures becomes significantly enhanced indicating a notable oxidation of the carbon structures in the sample -- as also already seen by XPS high resolution measurements. The oxidation process seems to be strongest for high magnetron powers and is probably related to the higher energy input into the sample. Since the magnetron is operated at higher powers than the RF-electrode  -- and since magnetrons are constructed in such a way that the plasma density is higher -- the amount as well as the energy of the ions hitting the substrate surface is much higher. This leads to an additional breaking of carbon bonds -- resulting in dangling bonds at the surface that saturate with atmospheric oxygen when brought in contact with air.

This is also confirmed by the information obtained by EDX and especially XPS -- the oxygen content of the cobalt containing samples is much higher than the one of pure polypyrrole.

Another feature -- which has not been detected for native pyrrole samples -- appears at 290.0\,eV and increases with raising magnetron power. In contrast to to the C=O peak it is already relatively intense for low magnetron power, but grows less pronounced when the magnetron power is increased. This points to a different origin of this peak. A comparison with previous studies suggests that this peak is related to the incorporation of Co into the sample and the formation of C-N-Co \cite{Schmidt2010} or C--Co \cite{Shao2009} bonds. Since XPS measurements have shown that only a small amount of carbon might be bound directly to cobalt, this peak is probably not mainly related to C--Co structures. In pyrrole rings a peak at around 289\,eV was assigned to several excitations -- C$_\alpha 1s$ to $3p\sigma, 3s\sigma, 3p\pi$ and $2\pi^*$. \cite{Duflot1998} However, it was also observed that this peak is higher in a cobalt containing porphyrin -- CoTPP -- then in a non-metal porphyrin -- 2HTPP. \cite{Schmidt2010} This was explained by complex orbital composition due to the cobalt $d_{xy}$ atomic orbital leading to such transitions with some Rydberg character.

Note that EDX has revealed that the relative Co content in the sample grows in the same fashion as the feature at 290.0\,eV. It is already high for 50\,W magnetron power and only grows slowly when the power is further raised (cf. Table \ref{tab:EDX}). Because the peak assigned to aromatic C-N structures in native pyrrole samples (at 286.6\,eV photon energy) almost vanishes with the appearance of the Co related peak at 290.0\,eV, we suspect that the Co binds to this chemical structure and that the 290.0\,eV peak mainly reflects newly formed C--N--Co structures. 

So it seems that C--N--Co structures similar to those present in CoTPP are formed in the dual plasma process. This is interesting because of the fact that those structures have been considered to be catalytically active \cite{Zagal2006}. Also note that both the C-N-Co assigned peak in the spectrum as well as the catalytic activity of the samples grow with magnetron power (cf. \cite{PartI}). A good way to check our proposition that catalytically active C--N--Co sites are formed in the plasma process is to look at the N $K$ absorption edge. If our assignment of the 290.0\,eV peak to C--N--Co structures is correct, an equivalent peak with a comparable dependence on the magnetron power should show up in the N $K$ edge spectrum, too.

\subsubsection{Nitrogen $K$ Edge}

The N $K$ edge spectra are shown in Figure \ref{fig:NEXAFS_N1s}. The spectrum obtained from the native polypyrrole sample shows 2 peaks related to the C--NH--C ring structure and the C=N--C bonds of polymerised pyrrole \cite{Schmidt2010, Hellgren2005}. At 401.3\,eV photon energy a third, little pronounced feature is detected. A similar near-edge structure has been also observed for pyrrole analogues. \cite{Mitra-Kirtley1993}

In the samples that were produced with the magnetron and thus contain cobalt a richer structure is observed. First of all, additional intensity builds up in the pre-edge region between $\sim389\,$eV and $\sim397\,$eV which grows as the magnetron power and thus the Co content in the sample is raised. This structure is not related to the nitrogen $K$ edge excitations. In fact, it reflects excitations originating from the Co $L_{2,3}$ edges at $\sim778\,$eV and $\sim793\,$eV due to second order light. Second order light has twice the photon energy of the one that is set at the monochromator and can be transmitted by the beamline to a small extent because it also fulfills the Bragg condition. Those Co $L_{2,3}$ edges are extremely intense and are located at almost twice the photon energy of the N $K$ edge excitations. Therefore, this second-order interference with the N $K$ edge is pronounced. We decreased the fix-focus constant $c_{ff}$ from the usual 2.25 down to 1.6 in order to reduce the second-order light transmission, yet a complete suppression could not be reached. However, the Co $L_{2,3}$ second-order contributions can easily be separated from the rest of the spectrum by comparison with the first-order signal measured at exactly twice the photon energy, as exemplarily shown for the 400\,W spectrum (grey shaded area).

Obviously, the second-order Co signal is not responsible for the peaks evolving at 401.3\,eV and 404.9\,eV photon energy which are increasing with magnetron power/Co content/catalytic activity. At the same time the peak at 399.5\,eV assigned to pyrrolic nitrogen seems to be decreasing in intensity, in agreement with what has been observed before at the C $K$ edge. 

Comparing this with previous experimental and theoretical studies, the two high-energy peaks can be attributed to several bindings that could be present within the compounds. For metal free catalysts -- so called carbon-alloys -- a peak at 401.5\,eV was observed and assigned to graphitic nitrogen, while a peak at 404.9\,eV was not observed. \cite{Niwa2009} However, both the feature at around 401.3\,eV and at 404.9\,eV were found for CoTPP -- a cobalt containing porphyrin \cite{Schmidt2010, Okajima2001}. We, therefore, assign the 404.9\,eV peak to Co related structures, namely the C--N--Co structure that has already been identified in the C $K$ edge spectra.

For the 401.3\,eV peak the situation is less clear. Previous studies on analogous compounds suggest that it could be either related to the above mentioned graphitic nitrogen or to amidic nitrogen (N--C=O). Since we have also found a pronounced C $K$ edge peak corresponding to C=O structures we would consider it more likely that the 401.3\,eV peak in our samples is related to amidic nitrogen.

\subsubsection{Oxygen $K$ Edge}

The oxygen $K$ edge spectra are shown in Figure \ref{fig:NEXAFS_O1s}. In native polypyrrole, small atmospheric oxygen contaminations are observed. In the magnetron-treated samples, additional structure evolves due to C=O and Co--O bonds \cite{Groot1993}. However, while the C=O signal is quite intense, the cobalt oxide feature remains rather small, which indicates that not all of the cobalt is not oxidised and other bonds, like Co--Co and Co--N are also produced (c.\,f. section~\ref{cha:NEXAFS:Co1s}).

\subsubsection{Cobalt $K$ Edge}
\label{cha:NEXAFS:Co1s}

The cobalt $K$ edge spectra are shown in figure \ref{fig:NEXAFS_Co1s} compared to cobalt oxide (CoO) and a pure metal cobalt foil. It can be seen that the spectra of cobalt oxide differs strongly from the other ones. So the amount of cobalt oxide within the samples is probably low. This correlates with the small Co--O peak that can be seen in the oxygen $K$ absorption edge and suggests that the fraction of cobalt oxide in the compounds is low. Comparing the cobalt--polypyrrol samples with the pure cobalt, it can be seen that the pre-peak at 7713\,eV increases with raising magnetron power. A peak at ca. 7710\,eV in pure cobalt is attributed to an absorption process $1s\rightarrow3d$. \cite{Moen1997} However, in a cobalt--phthalocyanine compound an additional structure at 7715\,eV was observed and attributed to a $1s\rightarrow4p_z$ transition related to a cobalt--nitrogen structure. \cite{Alves1992} Since the peak increases with rising magnetron power, the amount of cobalt--nitrogen structures seems to be higher at high magnetron powers. However, the cobalt--polypyrrole spectra resemble the spectrum of metallic cobalt very much so it seems that most of the cobalt within the samples is still bound in a metallic state. But since EDX measurements have shown that the cobalt content reaches 63\,at.\% while nitrogen content is only 5\,at.\%, it is obvious that not every cobalt atom can be part of the cobalt--nitrogen structures.

So also in the cobalt $K$ edge spectra, indications for the formation of Co--N structures similar to the ones that were observed in cobalt containing porphyrins are found. Therefore it seems clear that those structures are actually produced by using a dual plasma process to produce cobalt--polypyrrole compounds. Furthermore, the more power is used the more cobalt--nitrogen structures form leading to the increased catalytic activity that was reported in part I of this contribution. There are two possible reasons for that behaviour. The first one is that -- since the higher magnetron power leads to an increase of cobalt content within the samples -- nitrogen might more easily bind to the cobalt within the films. The second reason is the fact that increasing magnetron powers lead to a rising plasma density and thus to a higher energy input into the sample due to stronger bombardment with ions and atoms. Since in chemically produced samples a high energy input -- by heating of the substrate to elevated temperatures -- is needed to produce catalytically active structures as well, this higher electron density might play the same role as the elevated temperature in pyrolised compounds.

However it is also shown that some of the cobalt is still in metallic state -- only a fraction of cobalt atoms is bound to nitrogen. So by changing the plasma parameters in a way that even more of those structures are produced it might be possible to increase the catalytic activity of the films further.

\subsubsection{catalytic activity vs. peak intensities}

\begin{figure}[h]	
	\centering
	\includegraphics[width=\linewidth]{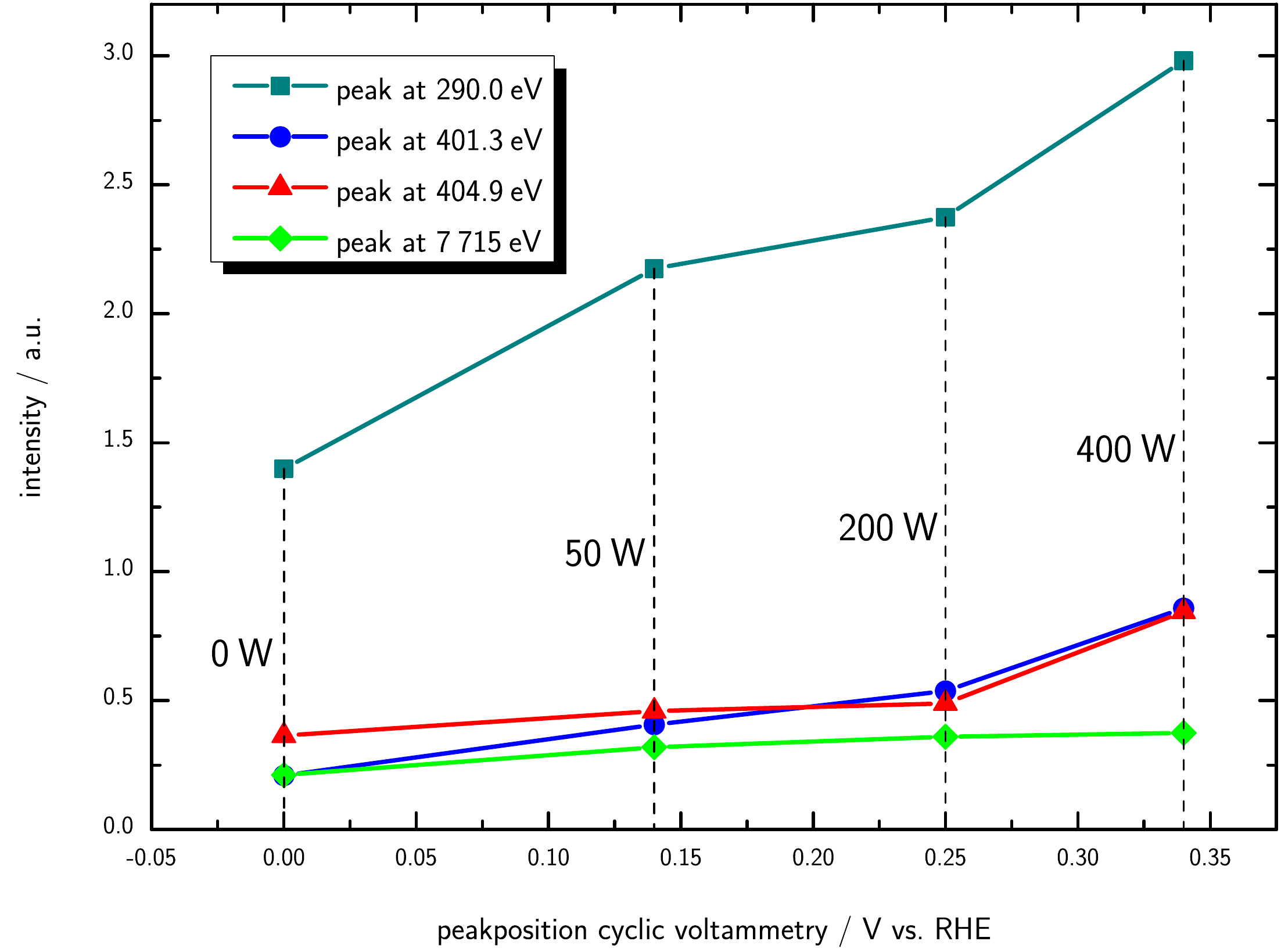} 
	\caption{correlation of peak position of the oxygen reduction peak in cyclic voltammetry with intensity of NEXAFS peaks; the power the samples were produced with is also given}
	\label{fig:correlation_VPR_Nexafs}
\end{figure}

In figure~\ref{fig:correlation_VPR_Nexafs} the peak position of the oxygen reduction reaction -- measured by cyclic voltammetry -- is plotted versus the intensities of the NEXAFS-peaks assigned to the Co--N--C structures. The magnetron power the samples were produced with is also marked in this figure. More information on the catalytic activity and measurement methods are given in part I of this contribution. \cite{PartI} 

It can easily be seen, that the intensity of the peaks, that can be correlated to the cobalt--nitrogen structures, is increasing with higher magnetron power and thus rising catalytic activity. This suggests that the Co--N--C structure is probably responsible for the catalytic activity of the films.

\subsection{Extended X-Ray absorption Fine Structure of Cobalt $K$ Edge}

\begin{figure}[h]
	\centering
	\includegraphics[width=\linewidth]{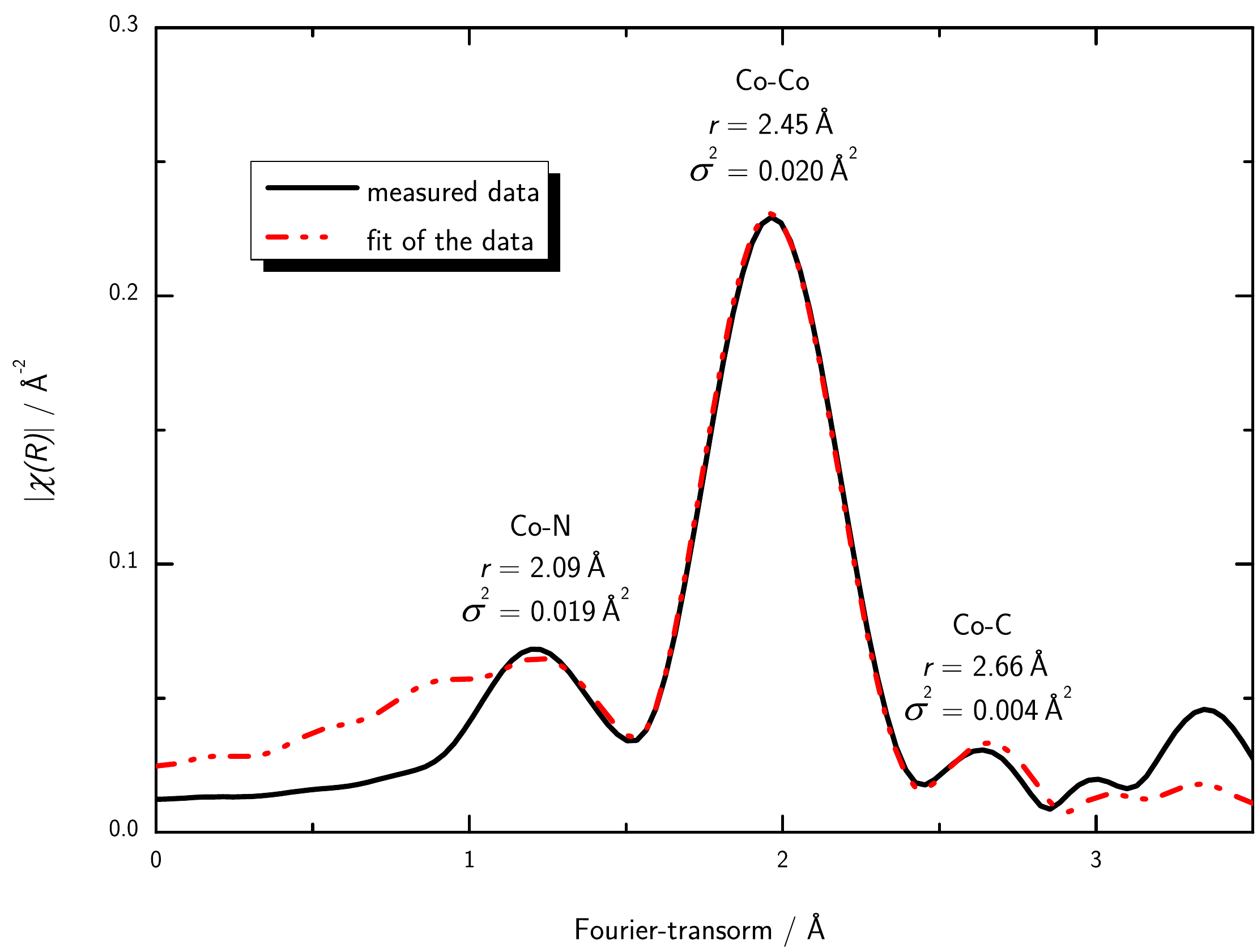} 
	\caption{Fourier-transform and fit of the $k^1$ weighted EXAFS data of the cobalt-polypyrrole sample produced with a magnetron power of 50\,W; the first shell data used for the fitting procedure and the values obtained are given next to the correspondig peak}
	\label{fig:EXAFS_Co1s}
\end{figure}

From the cobalt $K$ edges already presented in the previous section, EXAFS calculations were carried out. The Fourier-transform of those calculations is presented in figure \ref{fig:EXAFS_Co1s} for cobalt-polypyrrole produced with a magnetron power of 50\,W. The fit was carried out considering the 2 different cobalt structures -- metallic cobalt and cobalt--nitrogen structures -- that were observed by NEXAFS and XRD. It was carried out within the $k$-values of 1\,\AA$^{-1}$--11\,\AA$^{-1}$ and considering the $r$-value up to 3.6\,\AA. The results of that fit are also given in figure \ref{fig:EXAFS_Co1s}. It has to be noted that, in principle, Co--O-bonds can not be distinguished from Co--N-bonds in EXAFS-measurements. So the bond length calculated in this section migth also refer to the Co--O- instead of the Co--N-structures. However, NEXAFS measurements showed that most of the oxygen within the samples is bound to carbon. Furthermore, Co--N--C structures could be found in C, N and Co $K$-edges and should also be present in the Fourier-transform of the EXAFS measurements. Last but not least, it was demonstrated by the comparison of XPS to EDX elemental content that most of the oxygen is present at the surface of the films. Since EXAFS is a bulk-sensitive technique (information depth approx. 1\,\textmu m) the oxygen contaminants at the surface should only play a minor role. We therefore think it is reasonable to believe that the bonds in the EXAFS measurements belong to Co--N--C rather then Co--O--C. However, it can not completely be excluded that Co--O-bonds are also present and overlapping the cobalt--nitrogen-structures.

The calculated atom--atom distances are also given in this figure. It is important to note that those distances do not mean that a bond has to be present between those atoms -- only the physical presence is probed with EXAFS. However,from the bond lengths calculated it can be concluded that there is a bond between Co--Co and Co--N and no bond between Co and C. Furthermore, from the data obtained it can be concluded that there is a C-N bond present and since the multiple scattering with the path Co:N:C:Co is also included in the fitting, a C--N bond length of 1.44\,\AA\ can be calculated.

The Co--Co distance corresponds well to the one calculated with XRD, so this structure can be associated with the cobalt hcp nanoparticles present in the sample. The C--N bond length is also close to 1.38\,\AA, which was calculated for the equilibrium distance in pyrrole. \cite{Kofranek1992}

In cobalt containing porphyrins with a Co--N$_4$ structure cobalt--nitrogen bond lengths between 1.96\,\AA\ and 1.98\,\AA\ are observed, \cite{Scheidt1994,Zheng2007} in porphyrins where the cobalt atom is bound to more than 4 nitrogen atoms, the distance ranges from 1.98\,\AA--2.19\,\AA. \cite{Summers1994} In Co(N(CN)$_2$)$_2$ bond lengths of 2.10\,\AA\ are observed. \cite{Manson1998} Furthermore, in chemically produced cobalt--polypyrrole structures EXAFS measurements yielded cobalt--nitrogen distances between 2.04\,\AA\ and 2.07\,\AA. \cite{Yuasa2005} So the bond lengths observed within the plasma-produced cobalt--polypyrrole structures fit very well to the ones observed at chemically produced ones.

Unfortunately, due to the high cobalt content in samples produced with higher magnetron powers, and because of the fact already discussed in the NEXFAS section that most of the cobalt is still bound in a metallic way, EXAFS calculation of the cobalt--nitrogen bond length cannot be carried out at higher magnetron powers. The cobalt--cobalt bond is dominating those spectra. A shoulder arising from the cobalt--nitrogen bond can still be seen, but trying to fit this peak leads to a high error level in the bond length. So it cannot be evaluated if the bond length changes with magnetron power from the EXAFS measurements directly. But since the peak positions observed with NEXAFS do not change, the catalytically active structure is probably the same at all samples -- only the amount of this catalytically active structures varies with magnetron power.

\section{Conclusion}

The chemical structure of cobalt--polypyrrole compounds was analysed by various means. 

XRD measurements showed that the size of the cobalt nanoparticles within the compounds is 3\,nm and thus smaller than the sizes that are observed at pyrolised samples. Furthermore, only metallic cobalt with the low temperature phase was observed. This is a proof that the substrate was not heated during the plasma process.

With XPS and NEXAFS measurements it could be seen that by using the magnetron to introduce cobalt into the samples a stronger oxidation of the polypyrrole matrix takes place. It was attributed to the higher energy input into the substrate because of the higher electron density in the magnetron plasma. This leads to a stronger fractionation of the pyrrole ring so that more and more unsaturated bonds are present. Those bonds can then saturate when the compound is brought in contact with air.

Furthermore, it was shown that more and more catalytically active cobalt--nitrogen structures are produced with rising magnetron power. This is probably because of the fact that the bombardment of the substrate with ions and atoms increases with rising magnetron power. This leads to a higher energy input into the sample so that more of these structures can be produced. Since in chemically produced samples high temperatures -- and thus also high energy inputs --- have to be used to produce those catalytically active structures, this might be an analogue to the plasma-produced samples. 

EXAFS measurements showed that the cobalt--nitrogen bond length is 2.09\,\AA\ and the carbon--nitrogen bond length is 1.44\,\AA. This fits well to the distances observed in chemically produced cobalt--polypyrrole compounds. 

Increasing the amount of cobalt--nitrogen structures -- and thus the catalytic activity -- even more by tuning the plasma process parameters seems possible. Since the energy input into the substrate plays the most important role, tuning of the plasma parameters so that the energy input into the sample raises is the most promising method. This can for example be done by increasing the magnetron power above 400\,W. This is -- in principle -- possible. However, due to specifications of the magnetron we used, higher powers could not yet be achieved. Moreover, biasing the substrate is known to change the energy input into the substrate \cite{Maissel1965}. So, applying a bias to the samples could also have a positive effect on the production of cobalt--nitrogen structures. And to favour the production of cobalt--nitrogen bonds even more, the addition of different gases to the plasma process (e.g. N$_2$, H$_2$, NH$_3$) could be helpful.

\section{Acknowledgements}

The authors acknowledge the experimental support at BESSY II by Oleg Vilkov (Russian-German beamline) and Igor Zizak (KMC-2 beamline).

\newpage

\listoftables

\begin{table}[h]
	\centering
	\caption{Elemental contents of different cobalt-polypyrrole samples; acquired by EDX measurements}
	\begin{tabular}{llllll}
	\toprule  
	magnetron power & Co & N & O & C & N/C \\ 
	W & el.-\% & el.-\% & el.-\% & el.-\% &  \\
	\midrule
	0 & --- & 19.1 & 5.7 & 75.2 & 0.25 \\
	50 & 40.1 & 9.6 & 15.1 & 35.2 & 0.27 \\ 
	100 & 54.1 & 9.0 & 7.6 & 29.3 & 0.31 \\ 
	200 & 58.9 & 6.8 & 8.9 & 25.4 & 0.27 \\ 
	300 & 59.0 & 6.7 & 12.7 & 21.6 & 0.31 \\ 
	400 & 62.5 & 5.2 & 15.6 & 16.7 & 0.31 \\ 
	\bottomrule 
	\end{tabular} 
	\label{tab:EDX}
\end{table}

\begin{table}[h]
	\centering
	\caption{Elemental contents of different cobalt-polypyrrole samples; acquired by XPS measurements}
	\begin{tabular}{llllll}
	\toprule
	magnetron power & Co & N & O & C \\ 
	W & el.-\% & el.-\% & el.-\% & el.-\% \\
	\midrule
	0 & --- & 14.3 & 9.2 & 76.5\\
	50 & 17.8  & 3.3 & 29.9 & 48.7 \\ 
	100 & 22.4 & 2.6 & 31.1 & 43.6 \\ 
	200 & 23.2 & 1.5 & 34.2 & 40.6 \\ 
	300 & 24.6 & 0.8 & 35.6 & 38.5 \\ 
	\bottomrule 
	\end{tabular} 
	\label{tab:XPS}
\end{table}

\begin{table}[h]
	\centering
	\caption{Binding energies used for fitting of C $1s$ high resolution peaks} 
	\begin{tabular}{lll}
	\toprule 
	Sample & Binding energy & Reference \\ 
	\ & eV & \\
	\midrule 
	C$_\beta$, C--Co & 283.45 & \cite{Pfluger1984,Wesner1986} \\
	C$_\alpha$/C--C/C--H & 285.00 & \cite{Pfluger1984} \\ 
	C--N & 285.80 & \cite{Beamson1992} \\ 
	C--OH/C=N/C$\equiv$N/CN$^+$ & 286.55 &  \cite{Beamson1992,Malitesta1995}\\   
	N--C--O/C=O/C=N$^+$ & 287.80 & \cite{Beamson1992,Malitesta1995}\\
	COO & 288.83 & \cite{Beamson1992}\\
	Pyrrole-shake-up, CO$_3$ & 289.70 & \cite{Malitesta1995}\\
	\bottomrule
	\end{tabular}
	\label{tab:XPS_bindingenergy}
\end{table}

\begin{table*}[h]
	\centering
	\caption{Photon energy of NEXAFS Transitions}
	\begin{tabular}{lllll}
	\toprule
	edge & bond associated with transition & transition & transition energy & Reference \\
	& & & eV &  \\
	\midrule
	C & C=C & $1s\rightarrow \pi_2^*$ & 285.0 & \cite{Newbury1986,Zhang2009}\\
	C & C--N (Pyrrole ring) & $1s\rightarrow \pi_2^*$ & 286.6 & \cite{Newbury1986,Zhang2009}\\
	C & C=O & $1s\rightarrow \pi^*$ & 288.3 & \cite{Zhang2009,Bai2010}\\
	C & C--N--Co; C-Co & see text & 290.0 &  \cite{Schmidt2010,Shao2009} \\
	C & C--C; C=O & $1s\rightarrow \sigma^*$ shape res. & 297.3 & \cite{Newbury1986,Zhang2009} \\
	C & C=C, CO$_3$ & $1s\rightarrow \sigma^*$ & 300.4 & \cite{Newbury1986,Stohr1992} \\
	\midrule
	N & C=N--C (pyridinic) & $1s\rightarrow \pi^*$ & 398.6 &  \cite{Schmidt2010,Hellgren2005} \\
	N & C--NH--C (pyrrolic) & $1s\rightarrow \pi^*$ & 399.5 & \cite{Schmidt2010} \\
	N & N--C=O, graphitic N and $>$N-Co +$\pi$ & see text & 401.3 & \cite{Bai2010,Hellgren2005,Schmidt2010}\\
	N & C--N--Co structures & see text & 404.9 & \cite{Schmidt2010}\\
	\midrule
	O & Co--O Co3d & $1s\rightarrow \pi^*$ & 529.5 & \cite{Wu2005,Groot1993,Chen1997}\\
	O & O=O & $1s\rightarrow \pi^*$ & 531.2 & \cite{Hitchcock1980} \\
	O & C=O & $1s\rightarrow \pi^*$ & 532.8 & \cite{Zhang2009}\\
	O & OH & $1s\rightarrow \sigma^*$ & 539.4 & \cite{Zhang2009}\\
	O & Co--O Co4sp & $1s\rightarrow \sigma^*$ & 543.7 & \cite{Wu2005,Groot1993} \\
	\midrule
	Co & Co--Co & $1s\rightarrow 3d$ & 7710 & \cite{Moen1997} \\
	Co & Co--N & $1s\rightarrow4p_z$ & 7715 & \cite{Alves1992} \\
	\bottomrule
	\label{tab:NEXAFSenergies}
	\end{tabular}
\end{table*}

\clearpage

\listoffigures

\clearpage

\section{References}

\bibliographystyle{plainnat}

\begin{thebibliography}{53}
\providecommand{\natexlab}[1]{#1}
\providecommand{\url}[1]{\texttt{#1}}
\expandafter\ifx\csname urlstyle\endcsname\relax
  \providecommand{\doi}[1]{doi: #1}\else
  \providecommand{\doi}{doi: \begingroup \urlstyle{rm}\Url}\fi

\bibitem[IFE()]{IFEFFIT}
Ifeffit - {IFEFFIT}.
\newblock http://cars9.uchicago.edu/ifeffit/.
\newblock URL \url{http://cars9.uchicago.edu/ifeffit/}.

\bibitem[Alves et~al.(1992)Alves, Dodelet, Guay, Ladouceur, and
  Tourillon]{Alves1992}
M.~C.~Martins Alves, J.~P. Dodelet, D.~Guay, M.~Ladouceur, and G.~Tourillon.
\newblock Origin of the electrocatalytic properties for oxygen reduction of
  some heat-treated polyacrylonitrile and phthalocyanine cobalt compounds
  adsorbed on carbon black as probed by electrochemistry and x-ray absorption
  spectroscopy.
\newblock \emph{The Journal of Physical Chemistry}, 96\penalty0 (26):\penalty0
  10898--10905, December 1992.
\newblock \doi{10.1021/j100205a054}.
\newblock URL \url{http://dx.doi.org/10.1021/j100205a054}.

\bibitem[Bai et~al.(2010)Bai, Liu, Cook, Abbott, and Himpsel]{Bai2010}
Yiqun Bai, Xiaosong Liu, Peter Cook, Nicholas~L. Abbott, and F.~J. Himpsel.
\newblock Characterization of surfaces presenting covalently immobilized
  oligopeptides using {Near-Edge} x-ray absorption fine structure spectroscopy.
\newblock \emph{Langmuir}, 26\penalty0 (9):\penalty0 6464--6470, May 2010.
\newblock \doi{10.1021/la101101a}.
\newblock URL \url{http://dx.doi.org/10.1021/la101101a}.

\bibitem[Balazs et~al.(2007)Balazs, Hossain, Brombacher, Fortunato, Korner, and
  Hegemann]{Balazs2007}
{DJ} Balazs, {MM} Hossain, E~Brombacher, G~Fortunato, E~Korner, and D~Hegemann.
\newblock {Multi-Functional} nanocomposite plasma coatings - enabling new
  applications in biomaterials.
\newblock \emph{{PLASMA} {PROCESSES} {AND} {POLYMERS}}, 4:\penalty0 S380--S385,
  April 2007.
\newblock ISSN 1612-8850.
\newblock \doi{10.1002/ppap.200731004}.
\newblock URL
  \url{http://apps.isiknowledge.com/full_record.do?product=WOS&search_mode=GeneralSearch&qid=5&SID=W2Ib13eB4N5AhIa1dOM&page=1&doc=1}.

\bibitem[Beamson and Briggs(1992)]{Beamson1992}
G.~Beamson and D.~Briggs.
\newblock \emph{High Resolution {XPS} of Organic Polymers}.
\newblock John Wiley \& Sons, November 1992.
\newblock ISBN 0471935921.

\bibitem[Bogdanoff et~al.(2004)Bogdanoff, Herrmann, Hilgendorff, Dorbandt,
  Fiechter, and Tributsch]{Bogdanoff2004}
P~Bogdanoff, I~Herrmann, M~Hilgendorff, I~Dorbandt, S~Fiechter, and
  H~Tributsch.
\newblock Probing structural effects of pyrolysed {CoTMPP-based}
  electrocatalysts for oxygen reduction via new preparation strategies.
\newblock \emph{{JOURNAL} {OF} {NEW} {MATERIALS} {FOR} {ELECTROCHEMICAL}
  {SYSTEMS}}, 7\penalty0 (2):\penalty0 85--92, April 2004.
\newblock ISSN 1480-2422.

\bibitem[Chen(1997)]{Chen1997}
J.~G. Chen.
\newblock {NEXAFS} investigations of transition metal oxides, nitrides,
  carbides, sulfides and other interstitial compounds.
\newblock \emph{Surface Science Reports}, 30\penalty0 (1-3):\penalty0 1--152,
  1997.
\newblock ISSN 0167-5729.
\newblock \doi{16/S0167-5729(97)00011-3}.
\newblock URL
  \url{http://www.sciencedirect.com/science/article/pii/S0167572997000113}.

\bibitem[Daniel et~al.(2007)Daniel, Duguet, and Belmonte]{Daniel2007}
A.~Daniel, T.~Duguet, and T.~Belmonte.
\newblock Description of a hybrid {PECVD-PVD} process: Application to {Zn-Si-O}
  and {Ti-Si-O} composites thin films.
\newblock \emph{Applied Surface Science}, 253\penalty0 (24):\penalty0
  9323--9329, October 2007.
\newblock ISSN 0169-4332.
\newblock \doi{10.1016/j.apsusc.2007.05.063}.
\newblock URL
  \url{http://www.sciencedirect.com/science/article/B6THY-4NWCGPN-2/2/8578b0e82a8de7ff2e627d808678d47e}.

\bibitem[Despax and Flouttard(1989)]{Despax1989}
B.~Despax and J.~L. Flouttard.
\newblock Synthesis of gold---carbon composites by simultaneous sputtering and
  plasma polymerization of propane in r.f. capacitively coupled diode system
  (13.56 {MHz)}.
\newblock \emph{Thin Solid Films}, 168\penalty0 (1):\penalty0 81--88, January
  1989.
\newblock \doi{10.1016/0040-6090(89)90691-3}.
\newblock URL
  \url{http://www.sciencedirect.com/science/article/B6TW0-46X3R4Y-B/2/c0c261f17a690e5e433d9ecd843fd75c}.

\bibitem[Duflot et~al.(1998)Duflot, Hannay, Flament, and
  {Hubin-Franskin}]{Duflot1998}
D.~Duflot, C.~Hannay, {J.-P.} Flament, and {M.-J.} {Hubin-Franskin}.
\newblock Electronic excitation of gaseous pyrrole and pyrazole by inner-shell
  electron energy loss spectroscopy.
\newblock \emph{The Journal of Chemical Physics}, 109\penalty0 (13):\penalty0
  5308, 1998.
\newblock ISSN 00219606.
\newblock \doi{10.1063/1.477149}.
\newblock URL \url{http://link.aip.org/link/JCPSA6/v109/i13/p5308/s1&Agg=doi}.

\bibitem[Eggert(2005)]{Eggert2005}
Frank Eggert.
\newblock \emph{{Standardfreie Elektronenstrahl-Mikroanalyse (mit dem EDX im
  Rasterelektronenmikroskop) (German Edition)}}.
\newblock Books On Demand, 2005.
\newblock ISBN 3833425997.
\newblock URL
  \url{http://www.amazon.com/Standardfreie-Elektronenstrahl-Mikroanalyse-Rasterelektronenmikroskop-German-Edition/dp/3833425997}.

\bibitem[Faubert et~al.(1999)Faubert, C\^{o}t\'{e}, Dodelet, Lef\`{e}vre, and
  Bertrand]{Faubert1999}
G.~Faubert, R.~C\^{o}t\'{e}, J.~P. Dodelet, M.~Lef\`{e}vre, and P.~Bertrand.
\newblock Oxygen reduction catalysts for polymer electrolyte fuel cells from
  the pyrolysis of {FeII} acetate adsorbed on 3,4,9,10-perylenetetracarboxylic
  dianhydride.
\newblock \emph{Electrochimica Acta}, 44\penalty0 (15):\penalty0 2589--2603,
  1999.
\newblock ISSN 0013-4686.
\newblock \doi{10.1016/S0013-4686(98)00382-X}.
\newblock URL
  \url{http://www.sciencedirect.com/science/article/B6TG0-44N683F-9/2/28890bfe0ee80b6baaaebbdc0b2b0aaf}.

\bibitem[Garsuch et~al.(2008)Garsuch, Yang, Bonakdarpour, and
  Dahn]{Garsuch2008}
Arnd Garsuch, Ruizhi Yang, Arman Bonakdarpour, and {J.R.} Dahn.
\newblock The effect of boron doping into {Co-C-N} and {Fe-C-N}
  electrocatalysts on the oxygen reduction reaction.
\newblock \emph{Electrochimica Acta}, 53\penalty0 (5):\penalty0 2423--2429,
  January 2008.
\newblock \doi{10.1016/j.electacta.2007.10.014}.
\newblock URL
  \url{http://www.sciencedirect.com/science/article/B6TG0-4PYGW2S-1/1/60c103c460a9181cb45829b08ebd185d}.

\bibitem[Groot et~al.(1993)Groot, Abbate, Elp, Sawatzky, Ma, Chen, and
  Sette]{Groot1993}
F~M F~de Groot, M~Abbate, J~van Elp, G~A Sawatzky, Y~J Ma, C~T Chen, and
  F~Sette.
\newblock Oxygen 1s and cobalt 2p x-ray absorption of cobalt oxides.
\newblock \emph{Journal of Physics: Condensed Matter}, 5\penalty0
  (14):\penalty0 2277--2288, April 1993.
\newblock ISSN 0953-8984.
\newblock \doi{10.1088/0953-8984/5/14/023}.
\newblock URL \url{http://iopscience.iop.org/0953-8984/5/14/023}.

\bibitem[Hellgren et~al.(2005)Hellgren, Guo, Luo, S\r{a}the, Agui, Kashtanov,
  Nordgren, \r{A}gren, and Sundgren]{Hellgren2005}
Niklas Hellgren, Jinghua Guo, Yi~Luo, Conny S\r{a}the, Akane Agui, Stepan
  Kashtanov, Joseph Nordgren, Hans \r{A}gren, and {Jan-Eric} Sundgren.
\newblock Electronic structure of carbon nitride thin films studied by x-ray
  spectroscopy techniques.
\newblock \emph{Thin Solid Films}, 471\penalty0 (1-2):\penalty0 19--34, January
  2005.
\newblock ISSN 0040-6090.
\newblock \doi{16/j.tsf.2004.03.027}.
\newblock URL
  \url{http://www.sciencedirect.com/science/article/pii/S0040609004003803}.

\bibitem[Hitchcock and Brion(1980)]{Hitchcock1980}
A.~P. Hitchcock and C.~E. Brion.
\newblock K-shell excitation spectra of {CO}, n2 and o2.
\newblock \emph{Journal of Electron Spectroscopy and Related Phenomena},
  18\penalty0 (1):\penalty0 1--21, 1980.
\newblock ISSN 0368-2048.
\newblock \doi{10.1016/0368-2048(80)80001-6}.
\newblock URL
  \url{http://www.sciencedirect.com/science/article/B6TGC-44C939W-6T/2/a82e50422aae6a577a988a0e11c36b5e}.

\bibitem[Holleman and Wiberg(1995)]{Holleman1995}
Arnold~Fr. Holleman and Egon Wiberg.
\newblock \emph{Lehrbuch der Anorganischen Chemie}.
\newblock Gruyter, 101., verb. u. stark erw. a. edition, 1995.
\newblock ISBN 3110126419.

\bibitem[Jaouen et~al.(2009)Jaouen, Herranz, Lef\`{e}vre, Dodelet, Kramm,
  Herrmann, Bogdanoff, Maruyama, Nagaoka, Garsuch, Dahn, Olson, Pylypenko,
  Atanassov, and Ustinov]{Jaouen2009}
Fr\'{e}d\'{e}ric Jaouen, Juan Herranz, Michel Lef\`{e}vre, {Jean-Pol} Dodelet,
  Ulrike~I. Kramm, Iris Herrmann, Peter Bogdanoff, Jun Maruyama, Toru Nagaoka,
  Arnd Garsuch, Jeff~R. Dahn, Tim Olson, Svitlana Pylypenko, Plamen Atanassov,
  and Eugene~A. Ustinov.
\newblock {Cross-Laboratory} experimental study of {Non-Noble-Metal}
  electrocatalysts for the oxygen reduction reaction.
\newblock \emph{{ACS} Applied Materials \& Interfaces}, 1\penalty0
  (8):\penalty0 1623--1639, 2009.
\newblock \doi{10.1021/am900219g}.
\newblock URL \url{http://dx.doi.org/10.1021/am900219g}.

\bibitem[Kofranek et~al.(1992)Kofranek, Kovar, Karpfen, and
  Lischka]{Kofranek1992}
Manfred Kofranek, Tomas Kovar, Alfred Karpfen, and Hans Lischka.
\newblock Ab initio studies on heterocyclic conjugated polymers: Structure and
  vibrational spectra of pyrrole, oligopyrroles, and polypyrrole.
\newblock \emph{The Journal of Chemical Physics}, 96\penalty0 (6):\penalty0
  4464--4473, March 1992.
\newblock \doi{10.1063/1.462809}.
\newblock URL \url{http://link.aip.org/link/?JCP/96/4464/1}.

\bibitem[K\"{o}rner et~al.(2010)K\"{o}rner, Aguirre, Fortunato, Ritter,
  R\"{u}he, and Hegemann]{Korner2010}
Enrico K\"{o}rner, Myriam~H Aguirre, Giuseppino Fortunato, Axel Ritter,
  J\"{u}rgen R\"{u}he, and Dirk Hegemann.
\newblock Formation and distribution of silver nanoparticles in a functional
  plasma polymer matrix and related ag+ release properties.
\newblock \emph{Plasma Processes and Polymers}, 7\penalty0 (7):\penalty0
  619--625, July 2010.
\newblock ISSN 1612-8869.
\newblock \doi{10.1002/ppap.200900163}.
\newblock URL
  \url{http://onlinelibrary.wiley.com/doi/10.1002/ppap.200900163/abstract}.

\bibitem[K\"{o}rner et~al.(2011)K\"{o}rner, Rupper, L\"{u}bben, Ritter,
  R\"{u}he, and Hegemann]{Korner2011}
Enrico K\"{o}rner, Patrick Rupper, J\"{o}rn~F. L\"{u}bben, Axel Ritter,
  J\"{u}rgen R\"{u}he, and Dirk Hegemann.
\newblock Surface topography, morphology and functionality of silver containing
  plasma polymer nanocomposites.
\newblock \emph{Surface and Coatings Technology}, 205\penalty0 (8-9):\penalty0
  2978--2984, January 2011.
\newblock ISSN 0257-8972.
\newblock \doi{16/j.surfcoat.2010.11.001}.
\newblock URL
  \url{http://www.sciencedirect.com/science/article/pii/S0257897210011461}.

\bibitem[Maissel and Schaible(1965)]{Maissel1965}
L.~I. Maissel and P.~M. Schaible.
\newblock Thin films deposited by bias sputtering.
\newblock \emph{Journal of Applied Physics}, 36\penalty0 (1):\penalty0 237,
  1965.
\newblock ISSN 00218979.
\newblock \doi{10.1063/1.1713883}.
\newblock URL \url{http://dx.doi.org/10.1063/1.1713883}.

\bibitem[Maldonado and Stevenson(2005)]{Maldonado2005}
Stephen Maldonado and Keith~J. Stevenson.
\newblock Influence of nitrogen doping on oxygen reduction electrocatalysis at
  carbon nanofiber electrodes.
\newblock \emph{The Journal of Physical Chemistry B}, 109\penalty0
  (10):\penalty0 4707--4716, March 2005.
\newblock \doi{10.1021/jp044442z}.
\newblock URL \url{http://dx.doi.org/10.1021/jp044442z}.

\bibitem[Malitesta et~al.(1995)Malitesta, Losito, Sabbatini, and
  Zambonin]{Malitesta1995}
C.~Malitesta, I.~Losito, L.~Sabbatini, and P.~G. Zambonin.
\newblock New findings on polypyrrole chemical structure by {XPS} coupled to
  chemical derivatization labelling.
\newblock \emph{Journal of Electron Spectroscopy and Related Phenomena},
  76:\penalty0 629--634, December 1995.
\newblock \doi{10.1016/0368-2048(95)02438-7}.
\newblock URL
  \url{http://www.sciencedirect.com/science/article/B6TGC-3THGKFF-GD/2/2590dc8850a2da0377b3233ebc65324e}.

\bibitem[Manson et~al.(1998)Manson, Kmety, Huang, Lynn, Bendele, Pagola,
  Stephens, {Liable-Sands}, Rheingold, Epstein, and Miller]{Manson1998}
Jamie~L. Manson, Carmen~R. Kmety, Qing-zhen Huang, Jeffrey~W. Lynn, Goetz~M.
  Bendele, Silvina Pagola, Peter~W. Stephens, Louise~M. {Liable-Sands},
  Arnold~L. Rheingold, Arthur~J. Epstein, and Joel~S. Miller.
\newblock Structure and magnetic ordering of {MII[N(CN)2]2} {(M} = co,
  ni){\textdagger}.
\newblock \emph{Chemistry of Materials}, 10\penalty0 (9):\penalty0 2552--2560,
  1998.
\newblock \doi{10.1021/cm980321y}.
\newblock URL \url{http://dx.doi.org/10.1021/cm980321y}.

\bibitem[Matter and Ozkan(2006)]{Matter2006}
Paul~H. Matter and Umit~S. Ozkan.
\newblock Non-metal catalysts for dioxygen reduction in an acidic electrolyte.
\newblock \emph{Catalysis Letters}, 109\penalty0 (3-4):\penalty0 115--123, July
  2006.
\newblock ISSN {1011-372X}.
\newblock \doi{10.1007/s10562-006-0067-1}.
\newblock URL \url{http://www.springerlink.com/content/6243255n1u13374k/}.

\bibitem[{Mitra-Kirtley} et~al.(1993){Mitra-Kirtley}, Mullins, Van~Elp, George,
  Chen, and Cramer]{Mitra-Kirtley1993}
Sudipa {Mitra-Kirtley}, Oliver~C. Mullins, Jan Van~Elp, Simon~J. George, Jie
  Chen, and Stephen~P. Cramer.
\newblock Determination of the nitrogen chemical structures in petroleum
  asphaltenes using {XANES} spectroscopy.
\newblock \emph{Journal of the American Chemical Society}, 115\penalty0
  (1):\penalty0 252--258, January 1993.
\newblock \doi{10.1021/ja00054a036}.
\newblock URL \url{http://dx.doi.org/10.1021/ja00054a036}.

\bibitem[Moen et~al.(1997)Moen, Nicholson, Rnning, Lamble, Lee, and
  Emerich]{Moen1997}
Arild Moen, David~G. Nicholson, Magnus Rnning, Geraldine~M. Lamble, {Jyh-Fu}
  Lee, and Hermann Emerich.
\newblock {X-Ray} absorption spectroscopic study at the cobalt k-edge on the
  calcination and reduction of the microporous cobalt silicoaluminophosphate
  catalyst {CoSAPO-34}.
\newblock \emph{Journal of the Chemical Society, Faraday Transactions},
  93\penalty0 (22):\penalty0 4071--4077, 1997.
\newblock ISSN 09565000.
\newblock \doi{10.1039/a704488g}.
\newblock URL
  \url{http://pubs.rsc.org/en/content/articlelanding/1997/ft/a704488g}.

\bibitem[Newbury et~al.(1986)Newbury, Ishii, and Hitchcock]{Newbury1986}
D~C Newbury, I~Ishii, and A~P Hitchcock.
\newblock Inner shell electron-energy loss spectroscopy of some heterocyclic
  molecules.
\newblock \emph{Canadian Journal of Chemistry}, 64\penalty0 (6):\penalty0
  1145--1155, June 1986.
\newblock ISSN 0008-4042.
\newblock \doi{10.1139/v86-900}.
\newblock URL \url{http://www.nrcresearchpress.com/doi/abs/10.1139/v86-900}.

\bibitem[Niwa et~al.(2009)Niwa, Horiba, Harada, Oshima, Ikeda, Terakura, Ozaki,
  and Miyata]{Niwa2009}
Hideharu Niwa, Koji Horiba, Yoshihisa Harada, Masaharu Oshima, Takashi Ikeda,
  Kiyoyuki Terakura, Jun-ichi Ozaki, and Seizo Miyata.
\newblock X-ray absorption analysis of nitrogen contribution to oxygen
  reduction reaction in carbon alloy cathode catalysts for polymer electrolyte
  fuel cells.
\newblock \emph{Journal of Power Sources}, 187\penalty0 (1):\penalty0 93--97,
  February 2009.
\newblock ISSN 0378-7753.
\newblock \doi{16/j.jpowsour.2008.10.064}.
\newblock URL
  \url{http://www.sciencedirect.com/science/article/pii/S0378775308019770}.

\bibitem[Okajima et~al.(2001)Okajima, Yamamoto, Ouchi, and Seki]{Okajima2001}
T~Okajima, Y~Yamamoto, Y~Ouchi, and K~Seki.
\newblock {NEXAFS} spectra of metallotetraphenylporphyrins with adsorbed
  nitrogen monoxide.
\newblock \emph{Journal of Electron Spectroscopy and Related Phenomena},
  114-116:\penalty0 849--854, March 2001.
\newblock ISSN 0368-2048.
\newblock \doi{10.1016/S0368-2048(00)00268-1}.
\newblock URL
  \url{http://www.sciencedirect.com/science/article/B6TGC-42HNN78-4T/2/1f4123ace5fb79653b302418e7870f87}.

\bibitem[Ozaki et~al.(2006)Ozaki, Tanifuji, Kimura, Furuichi, and
  Oya]{Ozaki2006}
Jun-ichi Ozaki, Shin-ichi Tanifuji, Naofumi Kimura, Atsuya Furuichi, and Asao
  Oya.
\newblock Enhancement of oxygen reduction activity by carbonization of furan
  resin in the presence of phthalocyanines.
\newblock \emph{Carbon}, 44\penalty0 (7):\penalty0 1324--1326, June 2006.
\newblock ISSN 0008-6223.
\newblock \doi{16/j.carbon.2005.12.026}.
\newblock URL
  \url{http://www.sciencedirect.com/science/article/pii/S0008622306000224}.

\bibitem[Pfluger and Street(1984)]{Pfluger1984}
P.~Pfluger and G.~B. Street.
\newblock Chemical, electronic, and structural properties of conducting
  heterocyclic polymers: A view by {XPS}.
\newblock \emph{The Journal of Chemical Physics}, 80\penalty0 (1):\penalty0
  544--553, January 1984.
\newblock \doi{10.1063/1.446428}.
\newblock URL \url{http://link.aip.org/link/?JCP/80/544/1}.

\bibitem[Scheidt and {Turowska-Tyrk}(1994)]{Scheidt1994}
W.~Robert Scheidt and Ilona {Turowska-Tyrk}.
\newblock Crystal and molecular structure of {(Octaethylporphinato)cobalt(II).}
  comparison of the structures of {Four-Coordinate} {M(TPP)} and {M(OEP)}
  derivatives {(M} = {Fe-Cu).} use of area detector data.
\newblock \emph{Inorganic Chemistry}, 33\penalty0 (7):\penalty0 1314--1318,
  March 1994.
\newblock \doi{10.1021/ic00085a017}.
\newblock URL \url{http://dx.doi.org/10.1021/ic00085a017}.

\bibitem[Schmidt et~al.(2010)Schmidt, Fink, and Hieringer]{Schmidt2010}
Norman Schmidt, Rainer Fink, and Wolfgang Hieringer.
\newblock Assignment of near-edge x-ray absorption fine structure spectra of
  metalloporphyrins by means of time-dependent density-functional calculations.
\newblock \emph{The Journal of Chemical Physics}, 133\penalty0 (5):\penalty0
  054703, 2010.
\newblock ISSN 00219606.
\newblock \doi{10.1063/1.3435349}.
\newblock URL \url{http://link.aip.org/link/JCPSA6/v133/i5/p054703/s1&Agg=doi}.

\bibitem[Shao et~al.(2009)Shao, Kugler, Dadyburjor, Rykov, and Chen]{Shao2009}
Huifang Shao, Edwin~L. Kugler, Dady~B. Dadyburjor, Sergey~A. Rykov, and
  Jingguang~G. Chen.
\newblock Correlating {NEXAFS} characterization of {Co-W} and {Ni-W} bimetallic
  carbide catalysts with reactivity for dry reforming of methane.
\newblock \emph{Applied Catalysis A: General}, 356\penalty0 (1):\penalty0
  18--22, March 2009.
\newblock ISSN {0926-860X}.
\newblock \doi{16/j.apcata.2008.11.012}.
\newblock URL
  \url{http://www.sciencedirect.com/science/article/pii/S0926860X08006960}.

\bibitem[Sirk et~al.(2008)Sirk, Campbell, and Birss]{Sirk2008}
A.~H.~C. Sirk, S.~A. Campbell, and V.~I. Birss.
\newblock Effect of preparation conditions of {Sol--Gel-Derived}
  {Co--N--C-Based} catalysts on {ORR} activity in acidic solutions.
\newblock \emph{Journal of The Electrochemical Society}, 155\penalty0
  (6):\penalty0 B592--B601, June 2008.
\newblock \doi{10.1149/1.2900107}.
\newblock URL \url{http://link.aip.org/link/?JES/155/B592/1}.

\bibitem[St\"{o}hr(1992)]{Stohr1992}
Joachim St\"{o}hr.
\newblock \emph{{NEXAFS} Spectroscopy}.
\newblock Springer Berlin Heidelberg, 1st ed. 1992. corr. 2nd printing edition,
  August 1992.
\newblock ISBN 3540544224.

\bibitem[Summers et~al.(1994)Summers, Petersen, and Stolzenberg]{Summers1994}
Jack~S. Summers, Jeffrey~L. Petersen, and Alan~M. Stolzenberg.
\newblock Comparison of the structures of the {Five-Coordinate}
  {Cobalt(II)Pyridine}, {Five-Coordinate} {Cobalt(III)} methyl, and
  {Six-Coordinate} {Cobalt(III)} methyl pyridine complexes of
  octaethylporphyrin.
\newblock \emph{Journal of the American Chemical Society}, 116\penalty0
  (16):\penalty0 7189--7195, 1994.
\newblock \doi{10.1021/ja00095a022}.
\newblock URL \url{http://dx.doi.org/10.1021/ja00095a022}.

\bibitem[Vickerman(1997)]{Vickerman1997}
John~C. Vickerman.
\newblock \emph{Surface Analysis - The Principal Techniques}.
\newblock Wiley, 1 edition, July 1997.
\newblock ISBN 0471972924.

\bibitem[Walter et~al.(2009)Walter, Br\"{u}ser, Quade, and
  Weltmann]{Walter2009}
Christian Walter, Volker Br\"{u}ser, Antje Quade, and {Klaus-Dieter} Weltmann.
\newblock Structural investigations of composites produced from copper and
  polypyrrole with a dual {PVD/PE-CVD} process.
\newblock \emph{Plasma Processes and Polymers}, 6\penalty0 (12):\penalty0
  803--812, 2009.
\newblock \doi{10.1002/ppap.200900056}.
\newblock URL \url{http://dx.doi.org/10.1002/ppap.200900056}.

\bibitem[Walter et~al.(2012)Walter, Kummer, Vyalikhb, Br\"user, Quade, and
  Weltmann]{PartI}
Christian Walter, Kurt Kummer, Denis Vyalikhb, Volker Br\"user, Antje Quade,
  and Klaus-Dieter Weltmann.
\newblock Using a dual plasma process to produce catalysts for the oxygen
  reduction reaction in fuel cells -- part i: characterisation of the catalytic
  activity and surface structure.
\newblock \emph{Journal of The Electrochemical Society}, 159\penalty0
  (8):\penalty0 F494--F500, 2012.
\newblock \doi{10.1149/2.078208jes}.
\newblock URL \url{http://dx.doi.org/10.1149/2.078208jes}.

\bibitem[Wesner et~al.(1986)Wesner, Linden, and Bonzel]{Wesner1986}
D.~A. Wesner, G.~Linden, and H.~P. Bonzel.
\newblock Alkali promotion on cobalt: Surface analysis of the effects of
  potassium on carbon monoxide adsorption and {Fischer-Tropsch} reaction.
\newblock \emph{Applied Surface Science}, 26\penalty0 (3):\penalty0 335--356,
  September 1986.
\newblock ISSN 0169-4332.
\newblock \doi{16/0169-4332(86)90074-7}.
\newblock URL
  \url{http://www.sciencedirect.com/science/article/pii/0169433286900747}.

\bibitem[Williamson and Hall(1953)]{Williamson1953}
{G.K} Williamson and {W.H} Hall.
\newblock X-ray line broadening from filed aluminium and wolfram.
\newblock \emph{Acta Metallurgica}, 1\penalty0 (1):\penalty0 22--31, January
  1953.
\newblock ISSN 0001-6160.
\newblock \doi{16/0001-6160(53)90006-6}.
\newblock URL
  \url{http://www.sciencedirect.com/science/article/pii/0001616053900066}.

\bibitem[Wu et~al.(2005)Wu, Huang, Okamoto, Tanaka, Lin, Chou, Fujimori, and
  Chen]{Wu2005}
W.~B. Wu, D.~J. Huang, J.~Okamoto, A.~Tanaka, {H.-J.} Lin, F.~C. Chou,
  A.~Fujimori, and C.~T. Chen.
\newblock Orbital symmetry and electron correlation in {NaxCoO2}.
\newblock \emph{Physical Review Letters}, 94\penalty0 (14):\penalty0 146402,
  April 2005.
\newblock \doi{10.1103/PhysRevLett.94.146402}.
\newblock URL \url{http://link.aps.org/doi/10.1103/PhysRevLett.94.146402}.

\bibitem[Yang et~al.(2007{\natexlab{a}})Yang, Bonakdarpour, Easton,
  {Stoffyn-Egli}, and Dahn]{Yang2007a}
Ruizhi Yang, Arman Bonakdarpour, E.~Bradley Easton, P.~{Stoffyn-Egli}, and
  J.~R. Dahn.
\newblock {Co--C--N} oxygen reduction catalysts prepared by combinatorial
  magnetron sputter deposition.
\newblock \emph{Journal of The Electrochemical Society}, 154\penalty0
  (4):\penalty0 A275--A282, April 2007{\natexlab{a}}.
\newblock \doi{10.1149/1.2435670}.
\newblock URL \url{http://link.aip.org/link/?JES/154/A275/1}.

\bibitem[Yang et~al.(2007{\natexlab{b}})Yang, Stevens, Bonakdarpour, and
  Dahn]{Yang2007}
Ruizhi Yang, Krystal Stevens, Arman Bonakdarpour, and J.~R. Dahn.
\newblock Dependence of the activity of sputtered {Co--C--N} oxygen reduction
  electrocatalysts on {Heat-Treatment} temperature.
\newblock \emph{Journal of The Electrochemical Society}, 154\penalty0
  (9):\penalty0 B893--B901, 2007{\natexlab{b}}.
\newblock \doi{10.1149/1.2750444}.
\newblock URL \url{http://link.aip.org/link/?JES/154/B893/1}.

\bibitem[Yuasa et~al.(2005)Yuasa, Yamaguchi, Itsuki, Tanaka, Yamamoto, and
  Oyaizu]{Yuasa2005}
Makoto Yuasa, Aritomo Yamaguchi, Hisayuki Itsuki, Ken Tanaka, Masakuni
  Yamamoto, and Kenichi Oyaizu.
\newblock Modifying carbon particles with polypyrrole for adsorption of cobalt
  ions as electrocatatytic site for oxygen reduction.
\newblock \emph{Chemistry of Materials}, 17\penalty0 (17):\penalty0 4278--4281,
  2005.
\newblock \doi{10.1021/cm050958z}.
\newblock URL \url{http://dx.doi.org/10.1021/cm050958z}.

\bibitem[Zagal et~al.(2006)Zagal, Bedioui, and Dodelet]{Zagal2006}
Jos\'{e}~H. Zagal, Fethi Bedioui, and {Jean-Pol} Dodelet.
\newblock \emph{N4-macrocyclic metal complexes}.
\newblock Springer, 2006.
\newblock ISBN {038728429X}, 9780387284293.

\bibitem[Zhang et~al.(1997)Zhang, Wu, Pu, Zhang, Jin, Tong, Zhu, Cao, Zhu, and
  Cao]{Zhang1997a}
Jing Zhang, Mei~Zhen Wu, Tian~Shu Pu, Zheng~Yang Zhang, Ruo~Peng Jin, Zhi~Shen
  Tong, De~Zhang Zhu, De~Xin Cao, Fu~Ying Zhu, and Jian~Qing Cao.
\newblock Investigation of the plasma polymer deposited from pyrrole.
\newblock \emph{Thin Solid Films}, 307\penalty0 (1-2):\penalty0 14--20, October
  1997.
\newblock \doi{10.1016/S0040-6090(97)00271-X}.
\newblock URL
  \url{http://www.sciencedirect.com/science/article/B6TW0-3W0GGYK-5G/1/570e6c80b710463293a5005766eb6a9e}.

\bibitem[Zhang(2008)]{Zhang2008b}
Jiujun Zhang.
\newblock \emph{{PEM} Fuel Cell Electrocatalysts and Catalyst Layers:
  Fundamentals and Applications}.
\newblock Springer, Berlin, 1., st edition. edition, October 2008.
\newblock ISBN 1848009356.

\bibitem[Zhang et~al.(2009)Zhang, Carravetta, Plekan, Feyer, Richter, Coreno,
  and Prince]{Zhang2009}
Wenhua Zhang, Vincenzo Carravetta, Oksana Plekan, Vitaliy Feyer, Robert
  Richter, Marcello Coreno, and Kevin~C. Prince.
\newblock Electronic structure of aromatic amino acids studied by soft x-ray
  spectroscopy.
\newblock \emph{The Journal of Chemical Physics}, 131\penalty0 (3):\penalty0
  035103, 2009.
\newblock ISSN 00219606.
\newblock \doi{10.1063/1.3168393}.
\newblock URL \url{http://link.aip.org/link/JCPSA6/v131/i3/p035103/s1&Agg=doi}.

\bibitem[Zheng et~al.(2007)Zheng, Li, Shi, Gao, and Kadish]{Zheng2007}
Min Zheng, Fangfang Li, Zujin Shi, Xiang Gao, and Karl~M. Kadish.
\newblock Electrosynthesis and characterization of {1,2-Dibenzyl} c60: A
  revisit.
\newblock \emph{The Journal of Organic Chemistry}, 72\penalty0 (7):\penalty0
  2538--2542, March 2007.
\newblock \doi{10.1021/jo062486t}.
\newblock URL \url{http://dx.doi.org/10.1021/jo062486t}.

\end{thebibliography}

\end{document}